\documentclass[a4paper,12pt]{article}

\pdfoutput=1 % if your are submitting a pdflatex (i.e.\ if you have
\usepackage[hmargin=.7in,vmargin=1.1in]{geometry}
\usepackage{indentfirst}
\usepackage{amsfonts}
\usepackage{mathrsfs}
\usepackage{amsmath}
\usepackage{amssymb}
\usepackage{authblk}
\usepackage{cite}
\usepackage{xcolor}
\usepackage{mathtools}
\usepackage{tensor}
\usepackage{graphicx}% Include figure files
\usepackage{bm}% bold math
\usepackage{upgreek}
\usepackage{braket}
\usepackage{color,soul}
\usepackage{csquotes}
\usepackage{caption}
\usepackage{subcaption}
\usepackage[section]{placeins}

\usepackage[bookmarksnumbered=true,bookmarksopen=true]{hyperref}
 \hypersetup{colorlinks,
             linkcolor=[rgb]{0,0.3,0.6}, 
             citecolor=[rgb]{0,0.3,0.6}, 
             urlcolor=[rgb]{0,0.3,0.6}}

\newcommand\me{\mathrm{e}}

\newcommand\pp{\uppi}
\newcommand{\dif}{\mathrm{d}}

\DeclareMathOperator{\diag}{diag}
\newcommand*\abs[1]{{\left|#1\right|}}

\newcommand{\widesim}[2][2]{
  \mathrel{\overset{#2}{\scalebox{#1}[1]{$\sim$}}}
}

\begin{document}

\title{\Large\textbf{Regular black holes with improved energy conditions and their analogues in fluids}}
	
\author[a,b]{Chen Lan\thanks{stlanchen@yandex.ru}}
\author[a]{Yan-Gang Miao\thanks{Corresponding author, miaoyg@nankai.edu.cn}}
\author[a]{Yi-Xiong Zang\thanks{zangyx@mail.nankai.edu.cn}}

\affil[a]{\normalsize{\em School of Physics, Nankai University, 94 Weijin Road, Tianjin 300071, China}}	
\affil[b]{\normalsize{\em Department of Physics, Yantai University, 30 Qingquan Road, Yantai 264005, China}}

\date{ }		
	
\maketitle

%%%%%%%%%%%%%%%%%%%%%%%%%%%%%%
\begin{abstract}
On the premise of the importance of energy conditions for regular black holes,
we propose a method to remedy those models that break the dominant energy condition, 
e.g., the Bardeen and Hayward black holes.
We modify the metrics but ensure their regularity at the same time, so that the weak, null, and dominant energy conditions are satisfied, with the  exception of the strong energy condition.
Likewise, we prove a no-go theorem for conformally related regular black holes,  
which states that the four energy conditions can never be met in this class of black holes.
In order to seek evidences for distinguishing regular black holes 
from singular black holes,
we resort to analogue gravity and regard it as a tool to mimic realistic regular black holes in a
fluid. The equations of state for the fluid are solved via an asymptotic analysis associated with a numerical method,
which provides a modus operandi for experimental observations, 
in particular, the conditions under which one can simulate realistic regular black holes in the fluid.
\end{abstract}
%%%%%%%%%%%%%%%%%%%%%%%%%%%%%%
	
\maketitle

\tableofcontents
	
%%%%%%%%%%%%%%%%%%%%%%%%%%%%%%%%%%%%%%%%%%%%%%%%%%
\section{Introduction}
\label{sec:intr}
%%%%%%%%%%%%%%%%%%%%%%%%%%%%%%%%%%%%%%%%%%%%%%%%%%

As is well-known, Einstein's general relativity lacks~\cite{Frolov:2016pav} the ultraviolet (UV) completeness
that is reflected~\cite{Buchbinder:2021wzv} in the singular solutions of Einstein's equations at the classical level and in the non-renormalizability at the quantum level.
Regular black holes (RBHs) \cite{Ansoldi:2008jw,lamy2018theoretical} 
which have no curvature singularities at the centers %spring up in order to launch a 
challenge the UV incompleteness at the classical level.
This challenge originated from the change of vacuum \cite{Sakharov:1966aja,Gliner:1966} 
and was implemented through various approaches, such as the introduction of nonlinear matter \cite{Ayon-Beato:1998hmi}, 
the deformation of the commutative spaces \cite{Nicolini:2005vd}, 
the regularization of  singularities by quantum effects \cite{Koch:2014cqa,Bojowald:2020dkb},
and the assistance of alternative theories of gravity \cite{Berej:2006cc,Olmo:2015axa,Chamseddine:2016ktu}.
Meanwhile, the apparent differences that can be tested between RBHs and singular BHs (SBHs)  
have motivated numerous studies \cite{Myung:2007av,Amir:2016cen,Lan:2020fmn}.
Whether we can distinguish RBHs from SBHs by theoretical and experimental evidences is 
a critical point for research programs in the field of RBHs. 
Because RBHs are widely considered to be related to quantum physics,
the discovery of RBHs in the universe will certainly provide a new hope to search for quantum gravity.

There are  two ways to construct RBHs. The first 
starts with establishing BH metrics via certain mathematical rules \cite{Hayward:2005gi,Bronnikov:2006fu,Balart:2014cga} that guarantee the finite curvatures at BH centers, 
followed by enduing these metrics with physical meanings;  
e.g., the action of matter was provided, and then the theory of RBHs was established \cite{Fan:2016hvf}.
The second way draws support from physical theories or phenomena,  e.g.,
the existence of a finite length scale \cite{Nicolini:2005vd} or the asymptotic safety \cite{Koch:2014cqa}, 
involves the derivation the corrected metrics, which give rise to finite curvatures.

Nevertheless, among all the RBHs constructed in the above two ways, more than a few models break physical conditions or conjectures, 
in particular, 
the limited curvature conjecture \cite{markov1982limiting}, which states that the curvature invariants should be bounded by some universal value,
the weak energy condition (WEC) that is associated with the second law of BH mechanics \cite{Bekenstein:1973mi} 
or the dominant energy condition (DEC) that is related to the causal structure of spacetime~\cite{curiel2017primer}. 
The violation of these energy conditions motivates us to interpret RBHs from the perspective of quantum corrections.

If the violation is located inside regular black holes, the interpretation would be reasonable. The first reason is that an event horizon prevents any observer from observing a black-hole interior, and the second one is that some unknown effects that occur within event horizons or reachable microscopic scales may cause quantum corrections.
However, if the violation occurs outside horizons, the problem immediately arises. For instance, if the DEC is violated outside horizons, the causal structure of spacetime will be broken, indicating that the observer outside horizons will encounter chaotic causal phenomena. This is not acceptable from the perspective of physics.

For more information about the energy conditions of RBHs, we refer the reader to
Ref.\ \cite{Maeda:2021jdc}, where the energy conditions of four well-known regular black holes are reviewed; in particular, the DEC is taken into consideration as a criterion to determine whether a regular black hole is realistic. 
In brief, the weak, null, and dominant energy conditions are the primary prerequisites for us to construct realistic RBHs, where the strong energy condition is an exception.
Note that the violation of the DEC is pointed out for some RBHs in Ref.\ \cite{Maeda:2021jdc}; our aim is to further study how to achieve the recovery of the DEC for these RBHs.

Analogue gravity as a tool of gaining insights into general relativity has shown~\cite{Barcelo:2005fc} its significance, 
representing a great leap from passively waiting for signals from external galaxies to actively studying BHs in ground laboratories.
Among various manifestations of analogue gravity, acoustic BHs (ABHs) have not only a long history \cite{Unruh:1980cg}
but also an active status in current research \cite{Yuan:2021ets,Hod:2021qem,Ling:2021vgk}.
Following our previous work \cite{Lan:2021ayk}, in 
which we proposed a new method to construct acoustic regular BHs (ARBHs),
we explore ARBHs in terms of the energy conditions of their {\em astronomical counterparts} in the present work.
Meanwhile, the simulation strategy used in the present work is different from our previous one;  i.e., we adopt the approach proposed in Ref.\ \cite{deOliveira:2021edr}, where the singular Schwarzschild and Reissner-Nordstr\"om spacetimes can be simulated in fluids.
Our aim is to investigate realistic RBHs with the help of analogue gravity and
try to find apparent evidences or phenomena for distinguishing RBHs from SBHs.

The remainder of this paper is organized as follows. In Sec.\ \ref{sec:correct-RBHs}, we clarify what a realistic RBH means by discussing the energy conditions. 
In Sec.\ \ref{sec:remedy}, we propose a remedy to those RBHs that break the dominant energy condition,
including the Bardeen BH, the Hayward BH, and their extensions.
Sec.\ \ref{eq:no-go} is dedicated to conformally related RBHs, where we prove a no-go theorem under two general situations.
In Sec.\ \ref{sec:simulation}, we simulate a realistic RBH in a fluid by using the properties of flows, where
two specific models 
are discussed in terms of the asymptotic analysis associated with the numerical method.
Inspired by the locally polytropic behaviors in the equations of state (EoSs) in the above section, we address the question of whether it is possible to obtain an RBH that possesses a globally polytropic EoS in Sec.\ \ref{sec:polytropic}.
Secs.\ \ref{sec:cylinder} and \ref{sec:low-dimension} cover cylindrical RBHs and lower dimensional RBHs with polar symmetry, respectively.
The conclusions, along with future outlooks, are summarized in Sec.\ \ref{sec:concl}.
The appendices are dedicated to detailed discussions of the differential inequalities (Apps.\ \ref{app:diff-ineq} and \ref{app:local-diff-ineq}), the derivation of Eq.~(\ref{eq:dec-generic}) (App.\ \ref{app:dec-generic}), 
the regularity conditions of $n$-dimensional RBHs (App.\ \ref{app:high-rbh}), and the asymptotic analysis for solving nonlinear differential equations (App.\ \ref{app:asym-sol}).

%%%%%%%%%%%%%%%%%%%%%%%%%%%%%%%%%%%%%%%%%%%%%%%%%%
\section{Realistic regular black holes}
\label{sec:correct-RBHs}
%%%%%%%%%%%%%%%%%%%%%%%%%%%%%%%%%%%%%%%%%%%%%%%%%%

Let us first consider the simplest case of RBHs whose metrics are spherically symmetric and of the following form:
\begin{equation}
\label{eq:metric-regular}
    g_{\mu\nu} =\diag\{-f,f^{-1},\xi^2, \xi^2\sin^2\theta\},\qquad f:=1-\frac{2M \sigma(\xi, x_i)}{\xi},
\end{equation}
where $f$ is the shape function,
$\xi$ is the radial coordinate, and
$\sigma$ is dimensionless and may contain several parameters $x_i$, $i=1,\ldots,N$, such as mass and charge.
Moreover, these parameters must appear in $\sigma$ via the combinations
$\xi^{n_0} x_1^{n_1}\cdots x_N^{n_N}
$, which are also dimensionless.
If every combination includes a non-zero $n_0$, we can reduce one parameter and obtain $N-1$ independent dimensionless parameters by following the Buckingham $\pi$ theorem \cite{Sedov:1977sd}.

From the mathematical perspective, 
the regularity of curvature invariants at BH centers demands the limit $\sigma\sim O(\xi^n)$ with $n\ge 3$, 
and the asymptotic flatness requires the limit $\sigma\sim O(\xi^m)$ with $0\le m< 1$ \cite{Lan:2021ngq}. 
When $m=1$, $f$ may converge to a non-zero and non-unit constant, such that the spacetime is Ricci flat with $R=0$, at infinity.
The general properties of the shape function are illustrated in Ref.
 \cite{Bronnikov:2006fu}.

Given a Riemann tensor, one can construct 17 curvature invariants in total, which are called the Zakhary-Mcintosh (ZM) invariants \cite{Zakhary:1997acs}.  
For the metric Eq.\ \eqref{eq:metric-regular}, 
all 17 ZM invariants consist of the combinations of $\sigma$ and its first and second derivatives, i.e., $\sigma'$ and $\sigma''$. If certain conditions are further considered, e.g., $\sigma$ is a positive definite or monotonic function of the radial coordinate, one can show that $\sigma\sim O(\xi^n)$ with $n\ge 3$ and simultaneously guarantee that all the ZM invariants are finite at the center of the RBH described by Eq.\ \eqref{eq:metric-regular}.

%%%%%%%%%%%%%%%%%%%%%%%%%%%%%%%%%%%%%%%%%%%%
\subsection{Energy conditions}
%%%%%%%%%%%%%%%%%%%%%%%%%%%%%%%%%%%%%%%%%%%%
From the physical perspective, the constructed RBHs should not violate the weak, null, and dominant energy conditions, which play important roles \cite{Balart:2014jia,Rodrigues:2017yry}. 
The three energy conditions together with the strong energy condition can be formulated in three classes via different approaches:
{\em geometric}, {\em physical}, and {\em effective} ones \cite{Curiel:2014zba}. 
If the mechanism of constructing RBHs does not change the gravitational part of Einstein's equations, 
e.g., the gravitational field coupled with nonlinear electric fields or magnetic monopoles \cite{Ayon-Beato:1998hmi,Bronnikov:2000yz},
the three classes of definitions are equivalent.
In this situation, the energy-momentum tensor can be represented via Eq.~\eqref{eq:metric-regular}, and Einstein's equations read as follows:
\begin{equation}
    T^\mu_{\;\nu}\coloneqq\frac{1}{8\pp}
    G^\mu_{\;\nu}=
    \diag\left\{-\frac{M \sigma '}{4 \pp  \xi^2},
    -\frac{M \sigma '}{4 \pp \xi^2},
    -\frac{M \sigma ''}{8 \pp  \xi},
    -\frac{M \sigma ''}{8 \pp  \xi}\right\},
\end{equation}
where the prime denotes the derivative with respect to $\xi$.
Because $T^{t}_{\;t}=T^{\xi}_{\;\xi}$, there is no need \cite{Zaslavskii:2010qz} to distinguish 
the definitions of energy densities inside and outside the horizon; i.e., the energy density inside the horizon is the same as that outside the horizon for the metric we are considering.
We define the energy density $\epsilon$ and pressures $p_\xi$ and $p_\perp$ by the diagonal components of $T^\mu_{\;\nu}$:
\begin{equation}
    \epsilon\coloneqq\frac{M \sigma '}{4 \pp  \xi^2},\qquad
    p_\xi \coloneqq-\frac{M \sigma '}{4 \pp  \xi^2},
    \qquad
    p_\perp\coloneqq-\frac{M \sigma ''}{8 \pp  \xi}.
\end{equation}
Thus, the four energy conditions can be cast \cite{Kontou:2020bta} in terms of $\sigma$ and its derivatives:
\begin{equation}
\label{eq:ECDs}
    \begin{split}
    \text{WEC}&:\quad  \sigma '\ge 0\;\cup\; \xi \sigma ''\le 2 \sigma ',  \\
    \text{NEC}&:\quad  \xi \sigma ''\le 2 \sigma',  \\ 
     \text{SEC}&:\quad   \sigma ''\le 0\;\cup\; \xi \sigma ''\le 2 \sigma ', \\
     \text{DEC}&:\quad \sigma '\ge 0\;\cup\; -2 \sigma'\le \xi   \sigma ''\le 2 \sigma',
    \end{split}
\end{equation}
where WEC denotes the weak energy condition, NEC denotes the null energy condition, SEC denotes the strong energy condition, and 
DEC denotes the dominant energy condition.

It is not difficult  from the four energy conditions in Eq.\ \eqref{eq:ECDs} to find that the NEC, i.e., $\xi \sigma ''\leq 2 \sigma '$, must be maintained, 
otherwise the other three conditions will be broken.
In other words, $\xi \sigma ''\leq 2 \sigma '$ is an inequality that ensures the four energy conditions; 
furthermore, this differential inequality can be solved using the  Gr\"onwall-Bellman lemma \cite{Demidovich:2008ms}, and its solution reads $\sigma\le \sigma_0 \xi^3$, where $\sigma_0:=\lim_{\xi\to 0}\sigma/\xi^3$ is a positive constant (see App.\ \ref{app:diff-ineq} for details).
As a counterexample, we consider the  widely discussed model \cite{Balart:2014cga} with $\sigma = \exp[-q^2/(2 M \xi)]$, where $q$ represents the charge, 
in which this inequality is invalid because of $\sigma_0=0$.
Therefore, the model of Ref.~\cite{Balart:2014cga} is suggested to be ruled out from realistic RBHs, as are its extensions \cite{Ling:2021olm}, because matter generating such RBHs breaks the four energy conditions.

Next, the first inequality in the WEC and DEC, i.e., $\sigma'\ge 0$, provides a solution, i.e., $\sigma\ge 0$, under the boundary condition $\sigma(\xi)|_{\xi=0}=0$;
that is, $\sigma$ is a non-negative and monotonically increasing function of $\xi$. 
It is known that the non-minimal Wu-Yang monopole \cite{Balakin:2006gv,Balakin:2016mnn,Liu:2019pov,Rayimbaev:2021vsq,Jusufi:2020odz} is a counterexample because its $\sigma$ function, i.e.,
$\sigma=-\xi^3 \left(Q^2-2 M \xi\right)/[2 M \left(2 q Q^2 +\xi^4\right)]$, where $Q$ is a charge parameter, is not monotonic and not strictly positive either. 
Thus, the WEC of the Wu-Yang monopole is broken, as is the DEC.
In addition, the breaking of $\sigma'\ge 0$ may lead to other problems in the construction of RBHs. 
For instance, when $\sigma$ is bell shaped, i.e., $\sigma =4\exp\left(-q^2/\xi^2\right)-\exp\left(-2 q^2/\xi^2\right)$, the corresponding BH has two horizons and all the curvature invariants are finite, but the extreme horizon radius is the maximum of the horizons, and the temperature is divergent as the radial coordinate approaches the extreme horizon radius.

The SEC implies an attractive interaction due to the  Landau-Raychaudhuri equation \cite{Hawking:1973uf}; i.e., when the affine parameter increases,
the expansion scalar of a family of neighboring time-like geodesics decreases
because of the condition
$-(\epsilon+p_\xi+2p_\perp)\propto \sigma''\le 0$.
Therefore, the violation of the SEC leads to a repulsive interaction. However, this violation is nothing to be concerned about, because the SEC of an RBH must be broken~\cite{Zaslavskii:2010qz} near an RBH center.
Moreover, the zeros of the equation, i.e., $\sigma''(\xi_*)=0$, separate the spacetime into different types of interactions,
which will be discussed later with  concrete examples.

Compared with the other energy conditions, the DEC has its particularity reflected in the inequality $\xi\sigma''\ge-2\sigma'$, 
which can be visualized from the Ricci curvature $R\propto \xi \sigma ''+2 \sigma '$;
i.e., the negative Ricci curvature  violates~\cite{Bargueno:2020ais} the DEC. 
However, the differential inequality $\xi\sigma''\ge-2\sigma'$ gives the solution $\xi\sigma\ge 0$ under the boundary conditions $\sigma(\xi)|_{\xi=0}=0=\sigma'(\xi)|_{\xi=0}$. This solution is trivial and provides no more constraints to the $\sigma$ function.
In practice, one does not need to verify the four energy conditions.
If the DEC is valid, the WEC ad NEC are also valid. 
Therefore, checking the DEC is enough to guarantee the WEC and NEC. As for the SEC, we do not need to check it individually for RBHs, because it is valid beyond an RBH central region but invalid within an RBH central region because of a repulsive interaction. As a result, if the DEC is maintained for an RBH, the WEC and NEC are ensured automatically.

In summary, we list the requirements for a realistic RBH from the perspective of energy conditions. If an RBH with the metric Eq.\ \eqref{eq:metric-regular} is realistic, its $\sigma$-function has the following behaviors:
\begin{itemize}
    \item $\sigma$ is a non-negative and monotonically increasing function of $\xi$, where $\xi\in[0,\infty)$;
    \item $\sigma$ must be bounded by $\sigma_0 \xi^3$ from above, i.e., $\sigma\le \sigma_0 \xi^3$, where $\sigma_0:=\lim_{\xi\to0}\sigma/\xi^3$ and $\sigma_0$ must be positive.
\end{itemize}
These two conditions are necessary but not sufficient for an RBH to be realistic; see App.\ \ref{app:diff-ineq} for a detailed explanation.
In the next section, we show that some well-known examples, 
such as the Bardeen and Hayward BHs, comply with these two conditions, but their dominant energy conditions are broken. To solve this problem, 
we provide a phenomenological approach to restore their dominant energy conditions.

%%%%%%%%%%%%%%%%%%%%%%%%%%%%%%%%%%%%%%%%%%%%%%%%%%%%%%%%%%%%%%%%%%
\section{Remedy to regular black holes breaking dominant energy condition}
\label{sec:remedy}
%%%%%%%%%%%%%%%%%%%%%%%%%%%%%%%%%%%%%%%%%%%%%%%%%%%%%%%%%%%%%%%%%%

The problem of Hayward BHs depicted~\cite{Hayward:2005gi} by $\sigma=\xi^3/(\xi^3+q^3)$, where $q$ represents the charge, 
is the violation of the DEC in the region $\xi> 2^{1/3} q$,
even if this $\sigma$ satisfies the two items above.
The reason is explained in App.\ \ref{app:diff-ineq}.

As we mentioned above regarding the special status of the DEC, Hayward BHs also violate the WEC and NEC when $\xi> 2^{1/3} q$. For constructing a Hayward-like BH that ensures the DEC in $\xi\in[0,\infty)$, we propose the following $\sigma$ function: 
\begin{equation}
  \sigma=  \frac{M^{\mu-3}\xi^3}{\xi^{\mu }+q^\mu},\label{modhayward}
\end{equation}
where $M^{\mu-3}$ is introduced for balancing the dimension. The DEC requires $2<\mu \le (\sqrt{145}-7)/2\approx 2.52$, under which the Hayward-like BH given by Eq.~(\ref{modhayward}) would be realistic in the whole region of $\xi$, i.e., $\xi\in[0,\infty)$.
Alternatively, the dimensionless $\sigma$ can be established via a parameterization, i.e., $\sigma=(\xi/l)^3/[1+(\xi/l)^{\mu }]$,
where $l$ is a parameter with the length dimension. 

A similar procedure can be applied to the Bardeen BH, which gives rise to a Bardeen-like $\sigma$ function:
\begin{equation}
\label{eq:re-bardeen}
  \sigma=  \frac{M^{3\mu/2-3} \xi^3}{\left(\xi^{\mu }+q^{\mu }\right)^{3/2}}.
\end{equation}
For this model, the DEC gives rise to $4/3<\mu \le\left(\sqrt{113}-7\right)/2\approx 1.82$.

In fact, we can construct a general $\sigma$ function
\begin{equation}
\label{eq:generic-sigma}
\sigma=  \frac{M^{\mu\nu-3} \xi^3}{\left(\xi^{\mu }+q^{\mu }\right)^{\nu}},
\end{equation}
which satsfies the DEC
if the parameters $\mu$ and $\nu$ take the values in the following regions (see App.\ \ref{app:dec-generic} for the derivation):
\begin{subequations}
\label{eq:dec-generic}
\begin{eqnarray}
\frac{2}{\nu }<\mu \leq \frac{1}{2} \sqrt{\frac{49 \nu +96}{\nu }}-\frac{7}{2}
&&\quad \text{when}\quad \frac{2}{5}<\nu \leq 3;
\\
\frac{2}{\nu }<\mu \leq \frac{3}{\nu }
&&\quad \text{when}\quad \nu>3.
\end{eqnarray}
\end{subequations}
It is not difficult to verify that the RBHs described by Eq.\ \eqref{eq:generic-sigma} are realistic
because the numerator plays a decisive role, $\sigma\sim O(\xi^3)$ when $\xi\to 0$, and the asymptotic flatness is maintained~\cite{Lan:2021ngq} simultaneously, i.e., $f\to 1$, 
because the power of the denominator, $\mu\nu$, is greater than 2 when $\xi\to \infty$.

Nevertheless, we cannot remedy all the RBHs that break the DEC by simply
changing the power of radial coordinates. If $\sigma$ is not a rational function, for instance, $\sigma = \exp[-q^2/(2 M \xi)]$, 
this model cannot be repaired. On the other hand,
although the RBH obtained via quantum corrections, e.g., Refs.\ \cite{Koch:2014cqa,Bojowald:2020dkb},
can be remedied via the above phenomenological method, the remedied model will lose the original motivation of quantum corrections.
Let us take the RG-improved Schwarzschild BH \cite{Bonanno:2000ep} as an example, 
which is motivated by the theory of gravitational asymptotic safety \cite{Weinberg:1979udq,Reuter:1996cp}.
The shape function reads
\begin{equation}
f=1-\frac{2G(r) M}{r}, \qquad
G(r)=\frac{G_0 r^3}{r^3+\omega G_0 (r+\gamma G_0 M)},
\end{equation}
where $G(r)$ is the running Newton constant, which plays a similar role to  the $\sigma$-function, 
$G_0$ is identified with the experimentally observed value of Newton's constant, and
 $\omega$ and $\gamma$ are two positive parameters.
This RG-improved BH is regular from the perspective of finite curvatures, 
but it breaks the DEC because $G(r)$ violates $-2 G'\le r  G''$ 
beyond a certain value $r_0$, where $r_0$ is determined by a positive root of the algebraic equation
$$
-6 \gamma ^2 G_0^3 M^2 \omega +3 \gamma  G_0 M r^3-8 \gamma  G_0^2 M r \omega -3 G_0 r^2 \omega +r^4=0.
$$
Thus, according to our remedy used above, 
we change $r^3$ to $r^\mu$ and multiply by $M^{3-\mu}$ for balancing the dimension in the denominator of $G(r)$:
\begin{equation}
\label{eq:modi-newton}
\widetilde G(r)=\frac{G_0 r^3}{M^{3-\mu}r^\mu+\omega G_0 (r+\gamma G_0 M)},
\end{equation}
which reveals that the DEC requires
 $0\le \mu \le  (\sqrt{145}-7)/2\approx 2.52$.
However, such a modification loses the original motivation of the RG-improvement, 
which can be understood from 
the distance scale $\lambda$ that provides the relevant cutoff for the Newton constant.
 Using Eq.\ \eqref{eq:modi-newton} and the formula~\cite{Bonanno:2000ep} $\widetilde G(r) =G_0 \lambda ^2/(G_0 \omega +\lambda ^2)$,
 we obtain  
\begin{equation}
\lambda^2=\frac{G_0  \omega r^3 }{M^{3-\mu}r^{\mu }-r^3+G_0 \omega  \left(r+\gamma  G_0 M\right)},
\end{equation}
and give the asymptotic behaviors at zero and infinity, respectively:
\begin{equation}
\lambda^2\widesim{r\to 0} \frac{r^3}{\gamma  G_0 M},\qquad
\lambda^2\widesim{r\to \infty}  - G_0 \omega,
\end{equation}
where the second one violates the original asymptotic requirement, i.e., 
$\lambda\widesim{r\to \infty} r$.

In the next section, we demonstrate that the conformally related RBHs cannot be repaired either, by
proving a no-go theorem.

%%%%%%%%%%%%%%%%%%%%%%%%%%%%%%%%%%%%%%%%%%%%%%%%%%%%%%%%%%%
\section{No-go theorem for conformally related regular black holes}
\label{eq:no-go}
%%%%%%%%%%%%%%%%%%%%%%%%%%%%%%%%%%%%%%%%%%%%%%%%%%%%%%%%%
We discuss two classes of conformally related regular black holes: the conformally related Schwarzschild-type black holes and the 
astronomical counterparts of the ARBHs with the unit speed of sound.

%%%%%%%%%%%%%%%%%%%%%%%%%%%%%%%%%%%%%
\subsection{Conformally related Schwarzschild-type black holes}
\label{sec:conf-schwarzschild}
%%%%%%%%%%%%%%%%%%%%%%%%%%%%%%%%%%%%%

We claim that one cannot establish a scale factor $\Omega$
that regularizes the Schwarzschild BH and makes the metric satisfy the DEC at the same time.
To specify our statement, let us first express the metric of conformally related Schwarzschild BHs  \cite{Bambi:2016wdn}, 
\begin{equation}
\label{eq:conf-schwarzchild}
\widetilde g_{\mu\nu}=\Omega\, g_{\mu\nu},\qquad
g_{\mu\nu}=\diag\left\{
-\left(1-\frac{1}{\xi}\right),
\left(1-\frac{1}{\xi}\right)^{-1},
\xi^2,
\xi^2 \sin^2\theta
\right\},
\end{equation}
where the scale factor is set to be $\Omega=\exp [S(\xi)]>0$ and $2M=1$ is chosen for the discussions in this subsection.
The metric being regularized implies that the corresponding curvature invariants are finite in the whole spacetime,
particularly, at the BH center.
Next, instead of observing the Kretschmann scalar $K$,
we concentrate on the contraction of two Weyl tensors 
$W_{\mu\nu\alpha\beta}$ and $W^{\mu\nu\alpha\beta}$, where $W\coloneqq W_{\mu\nu\alpha\beta} W^{\mu\nu\alpha\beta}$,
which is referred to as the Weyl curvature hereinafter.
Because of the Ricci decomposition~\cite{Weinberg:1972kfs,Wald:1984rg}, we obtain $W=K - 2 R_2+R^2/3$ and see that 
the Kretschmann scalar and Weyl curvature are equivalent for diagnosing the singularity in the four-dimensional spacetime, 
where $R_2\coloneqq R_{\mu\nu} R^{\mu\nu}$ is the contraction of two Ricci tensors and $R\coloneqq g^{\mu\nu}R_{\mu\nu}$ is the Ricci scalar. The Weyl curvature corresponding to the metric Eq.\ \eqref{eq:conf-schwarzchild} reads
\begin{equation}
W=\frac{12\; \me^{-2 S(\xi)}}{\xi^6},
\end{equation}
which is finite at the BH center if  $\me^{- S(\xi)}$ converges to zero no slower than $\xi^3$.
When $\me^{- S(\xi)}$ converges to zero on the order of $\xi^3$, i.e., $\me^{- S(\xi)}\sim O(\xi^3)$, 
$S(\xi)$ {\em diverges positively},
and its first-order derivative must be negative.
In contrast, 
the asymptotic flatness requires $\Omega\to 1$ as $\xi\to \infty$;
i.e., $S(\xi)$ must converge to zero at infinity.
Summarizing the above properties of $S(\xi)$, we find that $\Omega^{-1}=\me^{-S}$ is a bounded function on the whole non-negative axis of $\xi$.

The energy conditions of the conformally related Schwarzschild BH given by Eq.\ \eqref{eq:conf-schwarzchild}
should be investigated inside and outside the horizon, 
because 
$T^t_{\; t}$ no longer equals $T^\xi_{\;\xi}$.
In other words, the constraint $T^t_{\; t}=T^\xi_{\;\xi}$ breaks the finiteness of the Weyl curvature.
This can be understood easily by solving $G^t_{\; t}=G^\xi_{\;\xi}$ as a differential equation of $S(\xi)$,
which provides a solution $\me^{-S}=c_2 (\xi+2 c_1)^2$ that converges to zero slower than $\xi^3$, where $c_1$ and $c_2$ are two integration constants.
Consequently, the energy conditions inside and outside the horizon are different 
and should  be treated separately.
The energy density and pressures are defined inside the horizon ($\xi<1$) as
\begin{equation}
\label{eq:energy-pressure-in}
\epsilon^{\rm in} \coloneqq-\frac{1}{8\pp} G^\xi_{\; \xi},\qquad
p_\xi^{\rm in} \coloneqq\frac{1}{8\pp} G^t_{\; t},\qquad
p_t^{\rm in} \coloneqq\frac{1}{8\pp} G^\theta_{\; \theta};
\end{equation}
and are defined outside the horizon ($\xi>1$) as
\begin{equation}
\label{eq:energy-pressure-out}
\epsilon^{\rm out} \coloneqq-\frac{1}{8\pp} G^t_{\; t},\qquad
p_{\xi}^{\rm out} \coloneqq\frac{1}{8\pp} G^\xi_{\; \xi},\qquad
p_t^{\rm out} \coloneqq\frac{1}{8\pp} G^\theta_{\; \theta},
\end{equation}
where $G^t_{\; t}$, $G^\xi_{\; \xi}$, and $G^\theta_{\; \theta}$
are components of the Einstein tensor calculated using the metric of Eq.\ \eqref{eq:conf-schwarzchild}.
Thus, the DEC is reduced to four differential inequalities in terms of 
$S(\xi)$ and its derivatives $S'(\xi)$ and $S''(\xi)$ in the range of $\xi<1$ or $\xi>1$. 
Among all the differential inequalities, 
$\epsilon^{\rm in}+p_\xi^{\rm in}\ge 0$ and 
$\epsilon^{\rm out}+p_\xi^{\rm out}\ge 0$ provide the same differential inequality:
\begin{equation}
\label{eq:key-conf-sch}
 (S')^2-2 S''\ge 0, \qquad 
\xi\in[0,1)\cup (1,\infty).
\end{equation}
Multiplying both sides of this inequality by a non-negative factor $\me^{-S/2}$, 
we arrive at
\begin{equation}
\me^{-S/2}\left[(S')^2-2 S''\right]
=\frac{\dif ^2}{\dif \xi^2} \left(4 \me^{-S/2}\right)\ge 0,
\end{equation}
from which we can conclude that $\me^{-S/2}$ is a convex function in the range of $\xi\in[0,1)\cup (1,\infty)$.
However, the finiteness of curvature invariants and asymptotic flatness of the metric demand that $\me^{-S}$ is
bounded; 
therefore, $\me^{-S/2}$ must be a constant,\footnote{If a differentiable and convex function is bounded on $\mathbb{R}$, it must be a constant, see e.g.\ Ref.\ \cite{Binmore:1982mat}.} 
which is obviously contradictory to the asymptotic behavior of the Weyl curvature at $\xi\to 0$, 
$\me^{-S}\sim O(\xi^n)$ with $n\ge 3$. 

In other words, there exists no such a conformal factor $\Omega$ 
that can regularize the Schwarzschild BH and guarantee the DEC simultaneously. 
This conclusion can be extended to the conformally related Schwarzschild-type BHs with singularity at $\xi=0$. For such a BH with the metric
\begin{equation}
\label{eq:metric-general}
 g_{\mu\nu}=\diag\{-f, f^{-1}, \xi^2, \xi^2 \sin^2\theta\}, \qquad
f=1-\frac{\sigma(\xi)}{\xi},
\end{equation}
where $\sigma(\xi)/\xi$ is of a unique pole at $\xi=0$ and goes to zero as $\xi\to\infty$, 
there exists no conformal factor $\Omega$ that satisfies the following two conditions simultaneously:
\begin{itemize}
\item The Weyl curvature of metric $\Omega\,   g_{\mu\nu}$ is finite in $\xi\in [0,\infty)$, where $\Omega\to 1$ at $\xi\to\infty$;
\item The DEC based on $\Omega\,  g_{\mu\nu}$ is valid. 
\end{itemize}
This is the so-called no-go theorem for the conformally related Schwarzschild-type BHs that belong to conformally related RBHs.

The proof exactly follows the case of conformally related Schwarzschild BHs.
At first, the Weyl curvature of metric Eq.\ \eqref{eq:metric-general} reads
\begin{equation}
W=\frac{\me^{-2 S} }{3 \xi^6}
\left[\xi \left(\xi \sigma ''-4 \sigma '\right)+6 \sigma \right]^2.
\end{equation}
Assuming that $\xi=0$ is the $d$-th order pole of $\sigma$, 
we can obtain an asymptotic relation: 
\begin{equation}
\label{eq:aympt-general}
\me^{- S} \sim O(\xi^{3+n}),\qquad n\ge d\ge 0,
\end{equation}
which ensures that the Weyl curvature is finite.
When $\xi\to\infty$, the asymptotic flatness demands $\me^{- S} \to 1$.
Moreover, the conditions $\epsilon^{\rm in}+p_\xi^{\rm in}\ge 0$ and 
$\epsilon^{\rm out}+p_\xi^{\rm out}\ge 0$ provide exactly the same differential inequality as Eq.\ \eqref{eq:key-conf-sch},
from which we find that $\me^{- S} $ is a convex function.
Nevertheless, the combination of the convexity and boundness of $\me^{- S} $ leads to a contradiction with the asymptotic relation of Eq.\ \eqref{eq:aympt-general}. Therefore, our statement is proved.

%%%%%%%%%%%%%%%%%%%%%%%%%%%%%%%%%%%%%%%%%%%%%%%%%%%%%%%%%%%%%%%%%%
\subsection{Astronomical counterparts of ARBHs with unit speed of sound}
%%%%%%%%%%%%%%%%%%%%%%%%%%%%%%%%%%%%%%%%%%%%%%%%%%%%%%%%%%%%%%%%%%

On the premise that the speed of sound is set to be unity,
we proposed~\cite{Lan:2021ayk} a general method to construct ARBHs in a fluid, 
where the metrics are similar to those of conformally related BHs \cite{Bambi:2016wdn}.
However, as we noted in Ref.~\cite{Lan:2021ayk}, the astronomical counterparts of ARBHs under a certain parameterization violate the DEC;
i.e., the ARBHs we constructed hardly have any physical counterpart in the universe.
Thus, it is natural to ask if we can find a way 
that 
the DEC for the astronomical counterparts of the ARBHs can be repaired and consequently the astronomical counterparts of the ARBHs can be detected in the universe.

The present case differs from that in the above subsection, 
as the unregularized metric $g_{\mu\nu}$ has no singularity.
Moreover, the ARBHs are different from the conformally related BHs because their conformal factors are proportional to the energy density of fluids and not constrained by any dynamical equations.

We discuss this in detail by following the strategy used in Ref.\ \cite{Lan:2021ayk}, which is opposite to that of Sec.\ \ref{sec:conf-schwarzschild}. We express \cite{Lan:2021ayk} the metric of {\em acoustic} RBHs with spherical symmetry:
\begin{equation}
\label{eq:metirc-conformal}
\widetilde g_{\mu\nu}=\rho
\diag\left\{
-f, f^{-1}, 
r^2, r^2 \sin^2\theta
\right\},\qquad f=1-v^2,
\end{equation}
where $r$ is the radial coordinate in {\em fluids},\footnote{We distinguish $r$ from $\xi$ --- the radial coordinate of {\em astronomical} BHs.} 
$\rho$ represents the mass density, and $v$ represents the radial velocity of fluids. The density and velocity are related~\cite{Lan:2021ayk} by
$v=A/(\rho r^2)$,
where $A$ is a positive constant. 
The radial velocity $v$ is supposed to be positive; otherwise, the density will be negative.
Moreover, because we have set the speed of sound to be unity, 
$v$ is dimensionless and $A$ has the same dimension as  $\rho r^2$.

By considering the similarity between Eqs.~\eqref{eq:conf-schwarzchild} and \eqref{eq:metirc-conformal} and dealing with $\rho$ as a scale factor, i.e., $\rho:=\me^{S(r)}$, we can express the Weyl curvature as
\begin{equation}
W=\frac{4 A^4 \me^{-6 S} }{3 r^{12}}
\left(2 r^2 S'^2-r^2 S''+10 r S'+15\right)^2,
\end{equation}
where the prime denotes the derivative with respect to $r$.
The regularity at $r=0$ demands
\begin{equation}
\label{eq:cond-reg-aco}
\me^{- S}\sim O(r^{n}),\qquad n\ge 2.
\end{equation}
Namely, $\rho$ diverges because $\rho\propto r^{-n}$, and $v$ converges to a constant~\cite{Lan:2021ayk} 
in order to make sure that the Weyl curvature is finite when $r\to 0$.
As mentioned in the second paragraph of this subsection, if $\rho$ was removed from Eq.\ \eqref{eq:metirc-conformal}, the remaining metric would still give rise to finite curvature invariants  everywhere,
which is different from the situation in Sec.\ \ref{sec:conf-schwarzschild}.
In addition, the asymptotic flatness requires $\rho\to 1$ or $v\sim O(r^{-2})$ as $r\to \infty$; i.e.,
$v$ is a monotonically decreasing function at infinity.

Similar to the discussion in Sec.\ \ref{sec:conf-schwarzschild}, we suppose that the acoustic metric Eq.~\eqref{eq:metirc-conformal} directly corresponds to a spacetime metric.
Thus, the energy density and pressures corresponding to the astronomical matter generating astronomical BHs must be defined inside and outside the horizon separately;
otherwise, the equation $G^{t}_{\;t}=G^{r}_{\;r}$ gives a false solution, i.e., $S(r)=c_4-2 \ln(r+2 c_3)$, where $c_3$ and $c_4$ are integration constants. If this solution
is consistent with the regularity Eq.\ \eqref{eq:cond-reg-aco}, we have $c_3=0$, which leads to the result that $f$ degenerates to a constant, i.e., $1-A^2c_4^2$, such that the corresponding metric is no longer a BH solution.
We then deduce that the forms of density and pressure inside the horizon must be different from those outside.
Next, following the proof process in Sec.\ \ref{sec:conf-schwarzschild}, we derive the inequality
from two similar inequalities, i.e.,
$\epsilon^{\rm in}+p_r^{\rm in}\ge 0$ and 
$\epsilon^{\rm out}+p_r^{\rm out}\ge 0$:
\begin{equation}
(S')^2- 2 S''\ge 0, \qquad
r\in \{r| r>0, \;v\ne 1\},
\end{equation}
which is similar to Eq.~(\ref{eq:key-conf-sch}). Therefore, we conclude that  $\me^{-S/2}$ is a convex function in the region of $r\in \{r| r>0, \;v\ne 1\}$, which
contradicts the regular condition of Eq.\ \eqref{eq:cond-reg-aco} and asymptotic flatness.
In other words, even if $\widetilde g_{\mu\nu}/\rho$ is regular in the sense of finite curvatures, 
the regularized metric $\widetilde g_{\mu\nu}$ cannot satisfy the DEC. That is, the DEC is violated in the astronomical counterparts of the ARBHs with the unit speed of sound.

%%%%%%%%%%%%%%%%%%%%%%%%%%%%%%%%%%%%%%%%%%%%%%
\section{Simulation of realistic RBHs in fluids}
\label{sec:simulation}
%%%%%%%%%%%%%%%%%%%%%%%%%%%%%%%%%%%%%%%%%%%%%%

Now, we discuss the simulation of realistic RBHs in a fluid.
Our aim is to construct the spacetime of realistic RBHs  with spherical symmetry using acoustic waves
and verify the conditions under which the realistic RBHs can be simulated in the fluid, i.e., find the equations of state. Our result may have guiding significance for experiments.

We start with the general stationary acoustic metric \cite{Barcelo:2005fc}
\begin{equation}
\label{eq:metric-general-fluid}
\dif s^2 =\frac{\rho}{c}\left[
-(c^2-v^2) \dif \tau^2 +\left(
\delta_{ij}+\frac{v^i v^j}{c^2-v^2}
\right) \dif x^i \dif x^j
\right],
\end{equation}
where $\rho$ and $v^i$ represent the mass density and velocity of fluids, respectively, 
$c\coloneqq \sqrt{|\partial p/\partial \rho|}$ represents the local speed of sound
and $p$ represents the pressure. 
We shall use Eq.\ \eqref{eq:metric-general-fluid} to simulate realistic RBHs
by providing the equations of state.

First, we suppose that the fluid is spherically symmetric and 
its velocity contains only a radial component, i.e., $v^i=\{v_r(r), 0, 0\}$; thus,
Eq.\ \eqref{eq:metric-general-fluid} is reduced to the following form in spherical coordinates: 
\begin{equation}
\label{eq:metric-original}
\dif s^2 =
\rho c \left[
-\left(1-\frac{v_r^2}{c^2}\right)\dif \tau^2
+\left(1-\frac{v_r^2}{c^2}\right)^{-1} \frac{\dif r^2}{c^2}
+\frac{r^2}{c^2}\dif \Omega^2
\right],
\end{equation}
where $\dif \Omega^2:= \dif \theta^2+\sin^2\theta\dif \phi^2$.
By using the solution of the continuity equation 
\begin{equation}
\label{eq:sol-continue}
\rho=\frac{A}{r^2 v_r},
\end{equation}
to replace $v_r$ and 
defining a new variable
\begin{equation} 
\xi^2 := \frac{r^2 \rho}{c},\label{newvar}
\end{equation}
we rewrite the acoustic metric as
\begin{equation}
\dif s^2 =
-F\dif \tau^2
+H  \dif \xi^2 
+\xi^2\dif \Omega^2,
\end{equation}
which is supposed to  equal the astronomical metric formally, where $F$ and $H$ are defined as follows:
\begin{equation}
\label{eq:met-comp}
F:=c \rho-\frac{A^2}{r^4 c \rho},\qquad
H:=\frac{4 r^4 c^4 \rho^4}{\left(r^4 c^2 \rho^2-A^2\right) 
\left[r \rho c'-c \left(r \rho '+2 \rho \right)\right]^2}.
\end{equation}
Here, the prime denotes the derivative with respect to $r$.
Note that our $F$ and $H$ are slightly different from those in Ref.\ \cite{deOliveira:2021edr}, where they were represented in terms of $\xi$ and $v$.
The purpose of our expression is to derive analytical expressions for the density $\rho$ and pressure $p$ of the fluid.

Second, we impose the condition $F H=1$, i.e., the simulated metric has only one shape function, which leads to the differential equation
\begin{equation}
\label{eq:constraint}
4 c^3 \rho^3=\left[r \rho  c'-c \left(r \rho '+2 \rho \right)\right]^2.
\end{equation}
Now, we solve $c$ using the first equation of Eq.\ \eqref{eq:met-comp}:
\begin{equation}
c=\frac{\beta}{\rho },\qquad
\beta\coloneqq 
\frac{F}{2}+\sqrt{\left(\frac{F}{2}\right)^2+\frac{A^2}{r^4}},\label{beta}
\end{equation}
where the negative root has been ignored owing to the positive $c$ and $\rho$.
By substituting $c=\beta/\rho$ into Eq.~\eqref{eq:constraint}, we obtain
\begin{equation}
\frac{\rho '}{\rho}=
-\frac{1}{r}+
\frac{\beta '}{2 \beta}
\pm\frac{\sqrt{\beta }}{r},
\end{equation}
where there are no rules for selecting any one of the two solutions at this moment. Then, we derive the density analytically:
\begin{equation}
\label{eq:density}
\rho_\pm
=\rho_0\frac{\sqrt{\beta}}{r} \exp\left(\pm
\int \frac{ \sqrt{\beta}}{r}\dif r
\right).
\end{equation}
where $\rho_0$ is an integration constant.
In addition, using Eqs.~(\ref{beta}) and (\ref{eq:density}) together with the definition of $c$, we compute the pressure:
\begin{equation}
\label{eq:pressure}
p_\pm=p_0-\frac{\beta^2}{\rho_\pm}+
	2\int \frac{\beta \beta'}{\rho_\pm} \dif r,
\end{equation}
where $p_0$ is an integration constant.
Eqs.\ \eqref{eq:density} and \eqref{eq:pressure} give the equation of state for the fluid.

In practice, $F$ as a function of $\xi$ corresponds to the shape function of the realistic RBH that we will to simulate in the fluid.
The relationship between $\xi$ and $r$, i.e., Eq.~(\ref{newvar}), can be represented \cite{deOliveira:2021edr} by 
the nonlinear differential equation
\begin{equation}
\label{eq:master}
A^2 [\xi(r)]^4+F[\xi(r)] r^6 [\xi(r)]^2 [\xi'(r)]^2-r^8 [\xi'(r)]^4=0,
\end{equation}
or equivalently by
\begin{equation}
\label{eq:master-relation}
A^2 \xi ^4 \left[\frac{\dif r(\xi)}{\dif \xi}\right]^4+F(\xi) \xi ^2 [r(\xi )]^6 \left[\frac{\dif r(\xi)}{\dif \xi}\right]^2-[r(\xi )]^8=0,
\end{equation}
which does not have analytical solutions generally.\footnote{We note that Eq.~(\ref{eq:master}) is more suitable for the numerical analysis for specific models in Sec.~\ref{sec:rembardeen} and Sec.~\ref{sec:rbhnonlinear}, while Eq.~(\ref{eq:master-relation}) is more suitable for the asymptotic analysis made below.} 
Furthermore, the variables associated with the fluid can be written in terms of the following functions of $r$:
\begin{equation}
\label{eq:phys-var}
c=\frac{r^2 \xi'}{\xi^2},\qquad
v=\frac{A}{r^2 \xi'},\qquad
\rho=\xi',\qquad
p'=\xi'' \left(\frac{r^2 \xi'}{\xi^2}\right)^2.
\end{equation}
We note from Eq.~(\ref{eq:phys-var}) that there exists a special position $r_c$ that is a stationary point of both $\rho$ and $p$,
and this point is determined by $\xi''(r_c)=0$ because $\rho'$ and $p'$ are proportional to $\xi''$. 
Detailed numerical analyses with concrete examples are presented below.

Before studying specific models,  we perform an asymptotic analysis for Eq.\ \eqref{eq:master-relation}
and provide general properties of solutions at $r\to 0$ and  $r\to\infty$.

For the simulated RBH at $r\to 0$, we obtain an asymptotic $F$ with the help of Ref.~\cite{Lan:2021ngq}, i.e.,
\begin{equation}
    F\sim 1-\frac{R(0)}{12} \xi^2,\label{Flimit} 
\end{equation}
when $\xi$ approaches $0$.
Here $R(0)$ is the limit of Ricci scalars at $\xi=0$.
Using the dominant balance \cite{Bender:2013amm,Paulsen:2013ap} and the boundary condition $\xi(r)|_{r=0}=0$, 
we find the asymptotic solution of Eq.\ \eqref{eq:master-relation} when $\xi$ approaches $0$ (see App.\ \ref{app:asym-sol} for details):
\begin{equation}
\label{eq:asymp-zero}
\xi \sim c_6\exp \left(-\frac{\sqrt{A}}{r}\right),
\end{equation}
where $c_6$ is an integration constant.
From Eqs.\ \eqref{beta}, \eqref{Flimit}, and \eqref{eq:asymp-zero}
we derive the asymptotic forms of $\beta$, the density,  and the pressure, respectively:
\begin{equation}
\label{eq:rho-p-fluid}
\beta\sim \frac{A}{r^2},\qquad
\rho_\pm \sim \rho _0\frac{\sqrt{A}  \me^{\mp \sqrt{A}/{r}}}{r^2},\qquad
p_\pm \sim -\frac{A^{3/2} \me^{\pm \sqrt{A}/{r}}}{r^2 \rho _0},
\end{equation}
which do not depend on specific RBH models at the leading order.
However, if we substitute Eq.\ \eqref{eq:asymp-zero} directly into Eq.\ \eqref{eq:phys-var},
we are able to fix $c_6$ and rule out the redundant root through comparison with Eq.\ \eqref{eq:rho-p-fluid}.
As a result, we obtain $c_6 = \rho_0$ and know that the solution with subscript ``$+$'' is physical; see App.\ \ref{app:asym-sol} for details.
Thus, we omit the subscript ``$+$'' in $\rho$  and $p$ for simplifying the notation in the following. 

We observe from Eq.~(\ref{eq:rho-p-fluid}) that the pressure of fluids at $r\to 0$ must be divergent, 
while the density 
converges to zero.
Furthermore, 
according to the first equation of Eq.\ \eqref{eq:phys-var}, we obtain the speed of sound:
\begin{equation}
c\sim \sqrt{A} \,\me^{\sqrt{A}/{r}},\label{speedlight}
\end{equation}
which is divergent at $r\to0$. 
In other words, if the maximum speed of sound exists \cite{trachenko2020speed}, 
there must be a cutoff $r_0$, such that the speed of sound is regularized.
This will be particularly important in the numerical calculation later.

Combining the density and pressure in Eq.\ \eqref{eq:rho-p-fluid}, 
we give the equation of state around the BH center: 
\begin{equation}
\label{eq:eos-zero}
p= -\frac{16}{\rho}  \left[W_{0}\left(-\frac{ \sqrt[4]{A}}{2} \sqrt{\frac{\rho }{\rho_0}}\right)\right]^4,
\end{equation}
where $W_0(\cdot)$ is the Lambert $W$ function.
To make Eq.\ \eqref{eq:eos-zero}
more intuitive, we perform the asymptotic expansion and give the leading order for $\rho\to 0$: 
\begin{equation}
p= -\frac{A \rho }{\rho_0^2}.
\end{equation}
Here, the relation between $\rho$ and $p$ around $r=0$ is linear.

In contrast,
for the simulated RBH at $r\to\infty$, we have $F\sim 1-2M/ \xi^{n}$, $0<n\le 1$ from Eq.~(\ref{eq:met-comp}), and
the asymptotic solution $\xi= c_7r$, where $c_7$ is constant (see App.\ \ref{app:asym-sol}). Thus,
we obtain $\beta\sim 1-2M/(c_7\, r)^n$ using Eq.~(\ref{beta}).
The density and pressure can be solved via Eqs.\ \eqref{eq:density} and \eqref{eq:pressure} with the plus subscript: 
\begin{equation}
\rho\sim \rho_0\, \me^{{M} /[n(c_7\, r)^{n}]},\qquad
p\sim p_0- \frac{(1-4 n) }{\rho _0} \me^{-{M }/[n(c_7\, r)^{n}]},
\end{equation}
where we have kept the leading term of $p$ valid by adding $n\neq 1/4$.
The limits of the density and pressure are
$\rho\to \rho_0$ and $p\to p_0-(1-4n)/\rho_0$, respectively. In particular,
the approximate equation of state at infinity reads
\begin{equation}
p= p_0-\frac{1-4 n}{\rho },
\end{equation}
which is polytropic; more precisely, it describes a  thermal process similar to that of the Chaplygin gas \cite{Kamenshchik:2001cp}.
If $n=1/4$, the equation of state becomes approximately
$p= p_0-1/(2 \rho)$.

Thus far, the discussions in this section have focused on the simulation of the RBHs given by Eq.~(\ref{eq:metric-regular}) in terms of acoustic analogy. In Secs.~\ref{sec:rembardeen} and \ref{sec:rbhnonlinear}, we simulate two realistic RBHs whose DEC is valid.
Details can be found in 
%as to the details of the two models, see 
Sec.~\ref{sec:remedy} for the first model and Ref.~\cite{Fan:2016hvf} for the second one.

%%%%%%%%%%%%%%%%%%%%%%%%%%%%%%%%%%%%%%%%%%%%
\subsection{Remedied Bardeen model}
\label{sec:rembardeen}
%%%%%%%%%%%%%%%%%%%%%%%%%%%%%%%%%%%%%%%%%%%%

\begin{figure}[!ht]
\centering
		\includegraphics[width=0.6\textwidth]{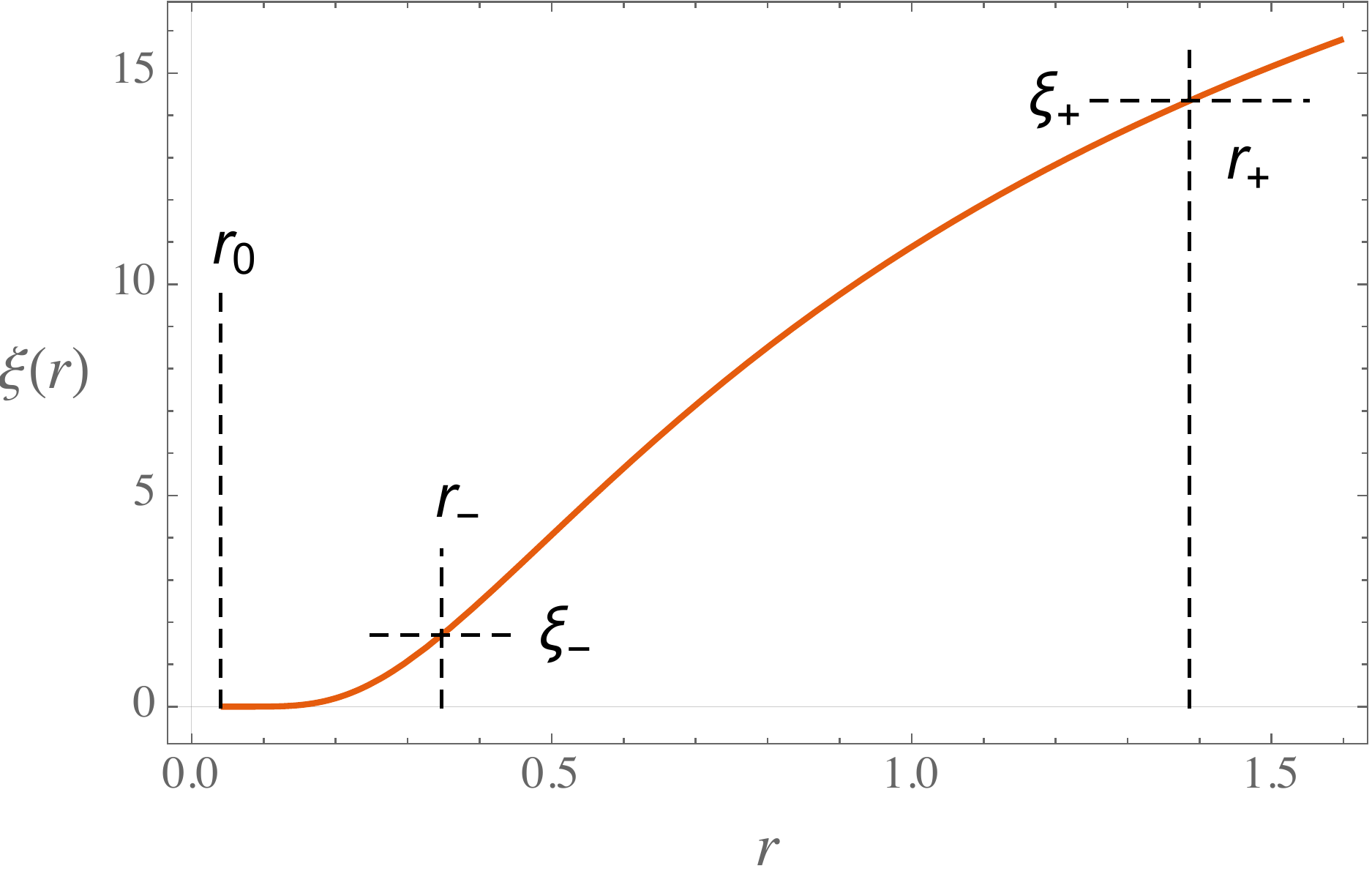}
		\caption{$\xi(r)$}
		\label{fig:Bardeen-xi}
\end{figure}

We simulate the remedied Bardeen BH by taking $\mu=3/2$, $M=1$, and $q=1$ in Eq.\ \eqref{eq:re-bardeen}.
The solution of Eq.\ \eqref{eq:master} obtained numerically with the boundary condition $\xi(0.2)=0.2$
is shown in Fig.\ \ref{fig:Bardeen-xi},
where the two horizons are $r_-\approx 0.35$ and $r_+ \approx 1.39$ or equivalently $\xi_-\approx1.69$
and 
 $\xi_+ \approx 14.34$ equivalently. 
The initial point starting at $r_0=0.2$ instead of $0$ is based on the possibility of the existence of the maximum speed of sound,
and such a setting can avoid dealing with the divergent speed of sound, velocity of fluid, and pressure in numerical calculations.
Furthermore, the critical point that separates the spacetime into different types of interactions is determined by
$\sigma''(\xi_*)=0$; see Eq.~(\ref{eq:ECDs}) and the following discussions about the SEC, i.e., $\xi_*=2^{2/3}\approx 1.59 < \xi_-$,
which is located inside the inner horizon. 

\begin{figure}[!ht]
	\centering
	\includegraphics[width=.6\textwidth]{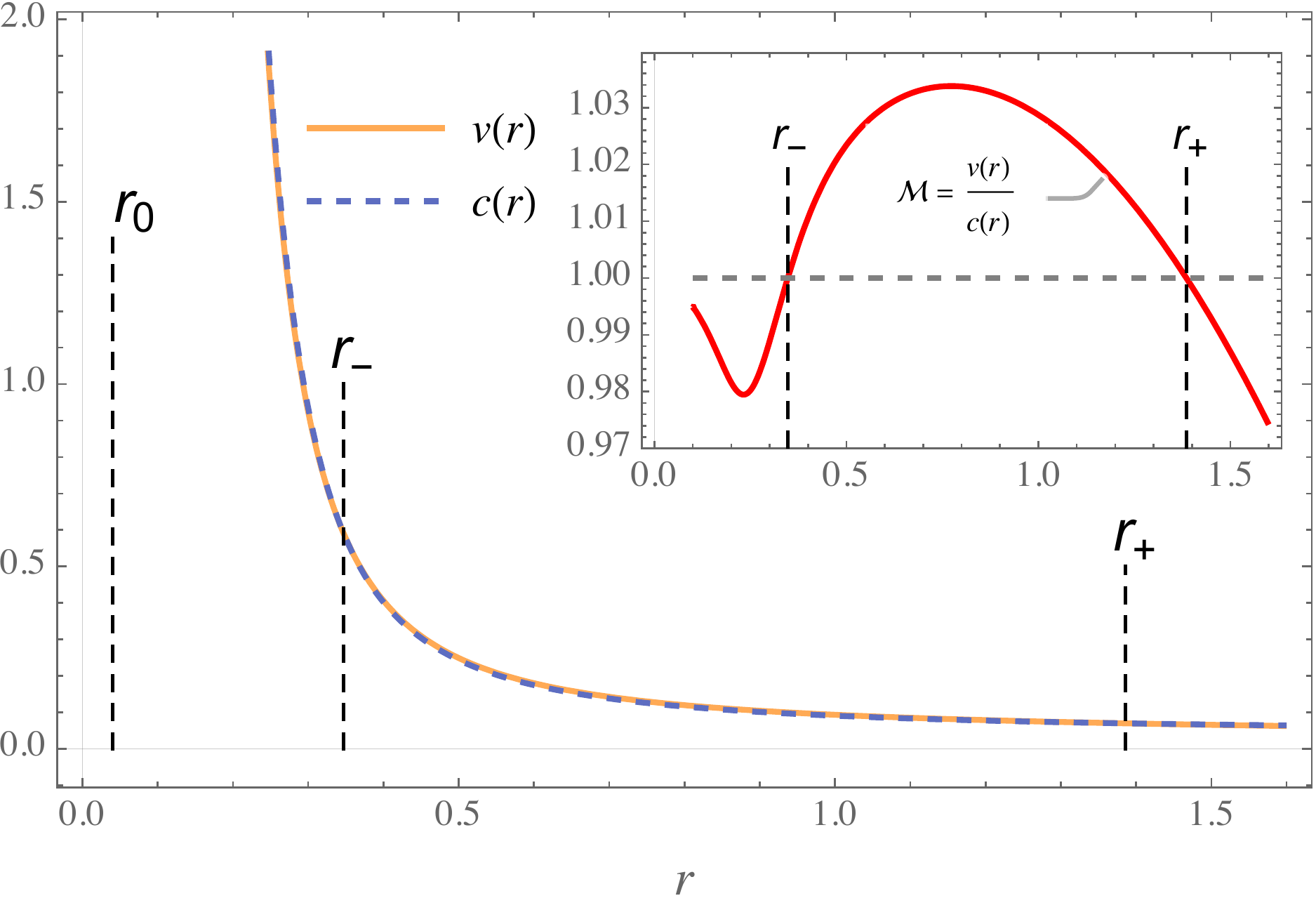}
	\caption{$v(r)$ and $c(r)$. The two curves are almost overlapped. For details, see the inset graph of $\mathcal{M}(r)$.}
	\label{fig:Bardeen-Mach}
\end{figure}

Fig.\ \ref{fig:Bardeen-Mach} shows the speed of sound and velocity of fluid and 
highlights their difference by using the Mach number, i.e., $\mathcal{M}\coloneqq v/c$.
We note that the Mach number is located in the range of $\mathcal{M}\in[0.8, 1.2]$ between the inner and outer horizons,
which indicates that the transonic phenomenon occurs. 
A similar phenomenon was observed~\cite{deOliveira:2021edr} for SBHs.
As a matter of fact, the existence of horizons for the acoustic model described by Eq.\ \eqref{eq:metric-original} 
separates the spacetime into different regions according to the signs of $c^2-v^2$. 
For the simulated RBHs with one horizon, the fluid inside the horizon flows with the transonic phenomenon.
For the simulated RBHs with two horizons, the transonic flow is sandwiched between the two horizons.

Generally, the Mach number can be computed with the help of Eq.\ \eqref{eq:phys-var}:
\begin{equation}
\mathcal{M}=\frac{A \xi^2}{r^4 (\xi')^2}
=\frac{1}{z+\sqrt{1+z^2}},\qquad z:=\xi^2F(\xi)/(2A).
\end{equation}
Because $F(\xi)<0$ as long as $\xi\in(\xi_-,\xi_+)$,
we have $z<0$ between the two horizons; meanwhile, 
we find that the Mach number is constrained by the following inequality:\footnote{
The function $\left(z+\sqrt{1+z^2}\right)^{-1}$ is positive and monotone decreasing because its derivative is negative,  $-1+z/\sqrt{z^2+1}<0$, and its limit at $z=0$ equals one.}
\begin{equation}
\label{eq:boundary-mach}
1<\mathcal{M}<\frac{1}{z_{\rm min}+\sqrt{1+z_{\rm min}^2}}.
\end{equation}
For the remedied Bardeen model, 
the minimum of $z$ can be calculated numerically, i.e., $z_{\rm min}\approx -0.033$,  under our setting, i.e., $A=M=q=1$, $\rho_0=1$, and $p_0=0$,
and this minimum corresponds to 
$r\approx 0.773$, which is located between the two horizons. Thus, the Mach number is in the range $1<\mathcal{M}<1.034$.

\begin{figure}[!ht]
\centering
	\begin{subfigure}{.478\textwidth}
	\centering
		\includegraphics[width=\textwidth]{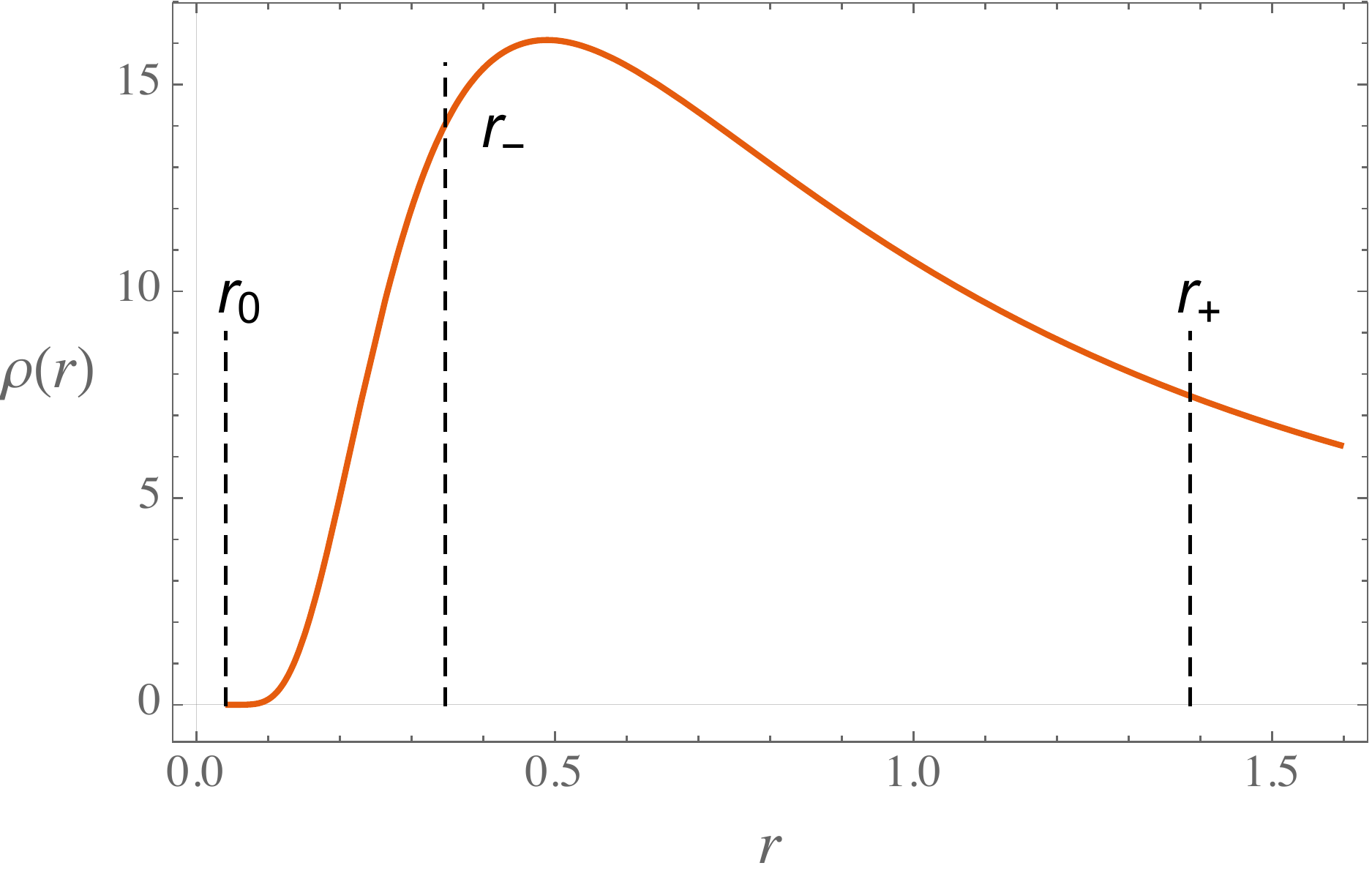}
		\caption{$\rho(r)$}
		\label{fig:Bardeen-rho}
	\end{subfigure}
	\begin{subfigure}{.502\textwidth}
	\centering
		\includegraphics[width=\textwidth]{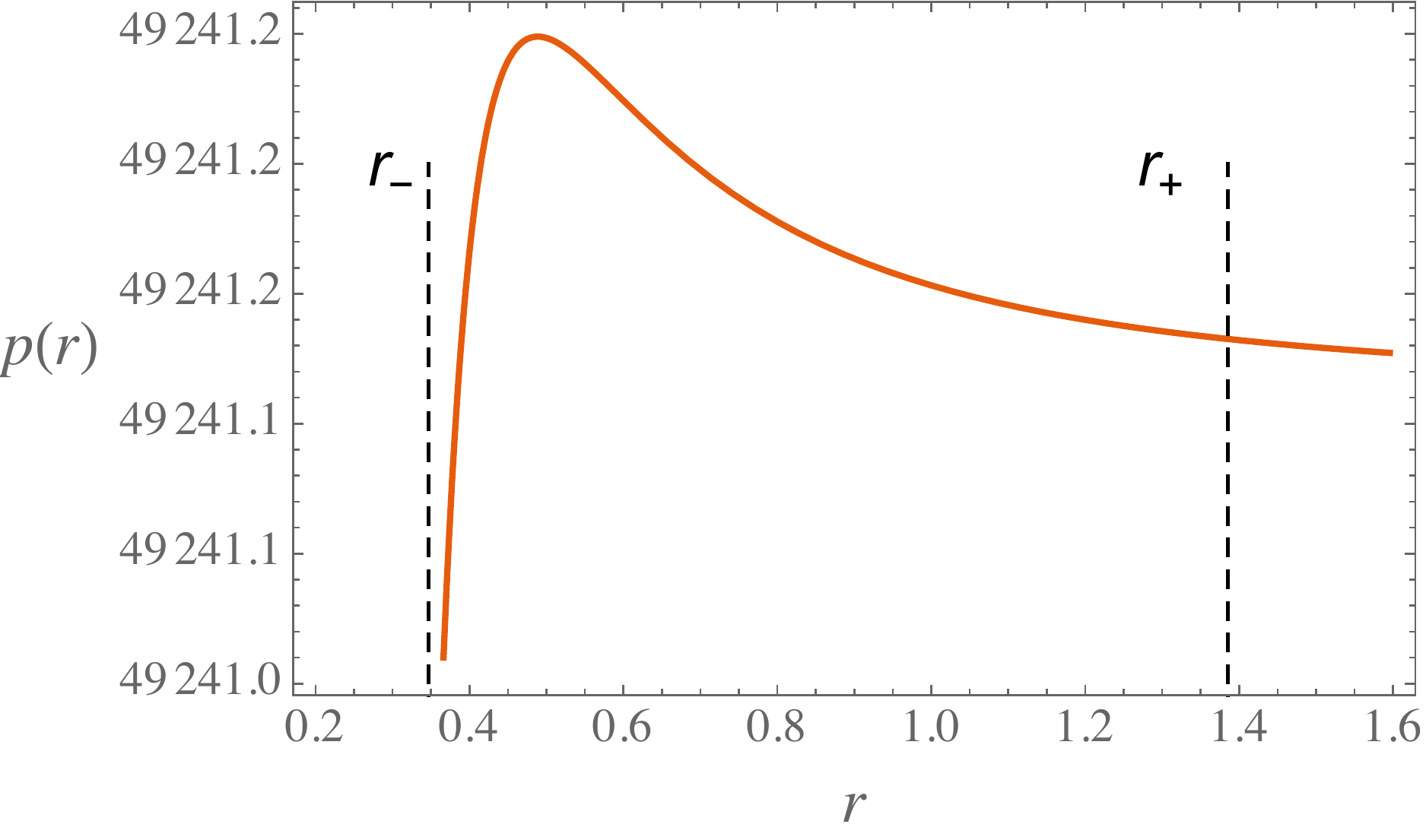}
		\caption{$p(r)$}
		\label{fig:Bardeen-p}
	\end{subfigure}
	\begin{subfigure}{.49\textwidth}
	\centering
		\includegraphics[width=\textwidth]{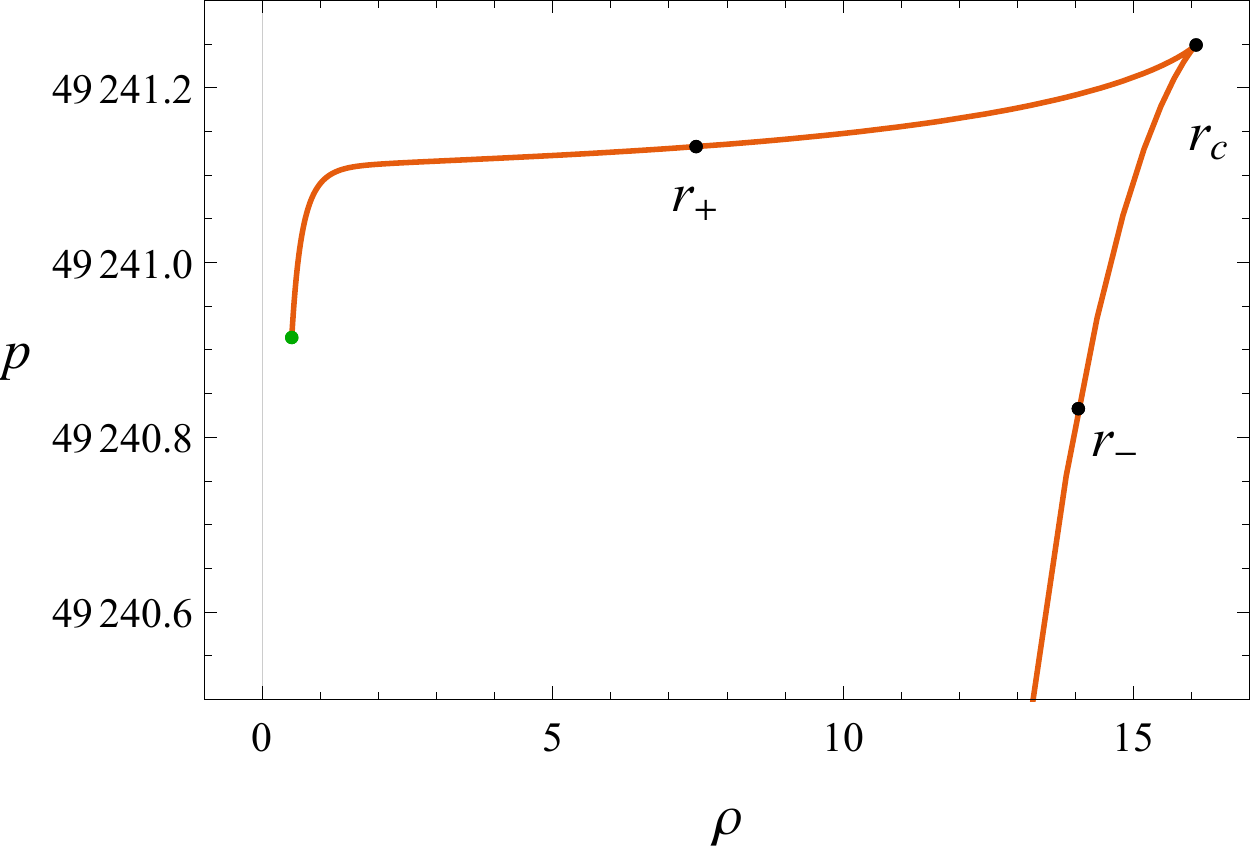}
		\caption{Equation of state}
		\label{fig:Bardeen-EoS}
	\end{subfigure}
	\caption{Numerical solutions for the remedied Bardeen model, where $A=M=q=1$, $\rho_0=1$, and $p_0=0$.}
	\label{fig:re-bardeen}
\end{figure}

We also provide the numerical calculations of Eqs.~\eqref{beta}, \eqref{eq:density}, and \eqref{eq:pressure} with subscript ``+'' for the density and  
pressure in Figs.\ \ref{fig:Bardeen-rho} and \ref{fig:Bardeen-p}, respectively, 
and the EoS in Fig.\ \ref{fig:Bardeen-EoS}. In Figs.\ \ref{fig:Bardeen-rho} and \ref{fig:Bardeen-p}, 
there are a global maximum of $\rho$ and a global maximum of $p$ located between the two horizons,   i.e., $r_c=0.489$. 
This point plays a special role in  Fig.\ \ref{fig:Bardeen-EoS} because it
shows a sharp discontinuity of the EoS.
In addition, the curve of the EoS ends at the green dot, where its values of $\rho$ and $p$, i.e., $\rho\approx0.509$ and $p\approx 49240.914$,  are estimated numerically when $r$ approaches $500$ as the infinity of our numerical calculations.

%%%%%%%%%%%%%%%%%%%%%%%%%%%%%%%%%%%%%%%%%%%%
\subsection{RBHs associated with nonlinear electromagnetic fields}
\label{sec:rbhnonlinear}
%%%%%%%%%%%%%%%%%%%%%%%%%%%%%%%%%%%%%%%%%%%%

Now, we turn to the model associated with nonlinear electromagnetic fields~\cite{Fan:2016hvf},
whose $\sigma$ is a rational and sigmoid function:
\begin{equation}
    \sigma=\frac{\xi^3}{(\xi+q)^3} \quad \text{with} \quad q\ge 0,
\end{equation}
where $\xi$ denotes the radial coordinate in the RBH. 
Note that $\sigma$ is non-negative and monotonically increasing in the whole region of $\xi$, i.e., 
$\sigma\ge 0$ and $\sigma'=q \xi^2/(q+\xi)^2\ge 0$;  
meanwhile, $\sigma$ is bounded by $\sigma\le \xi^3/q^3$ because of $\sigma_0=1/q^3$.
The critical point of this model can be obtained by solving $\sigma''(\xi_*)=0$,
which gives $\xi_*=q$.
Moreover, the existence of horizons demands $q\le 4 \xi_{\rm Sch} /27 $; i.e.,
the critical point $\xi_*$ is 
not greater than $4 \xi_{\rm Sch} /27$, where $\xi_{\rm Sch}=2M$ is the Schwarzschild horizon radius.
 
\begin{figure}[!ht]
\centering
	\begin{subfigure}{.49\textwidth}
	\centering
		\includegraphics[width=\textwidth]{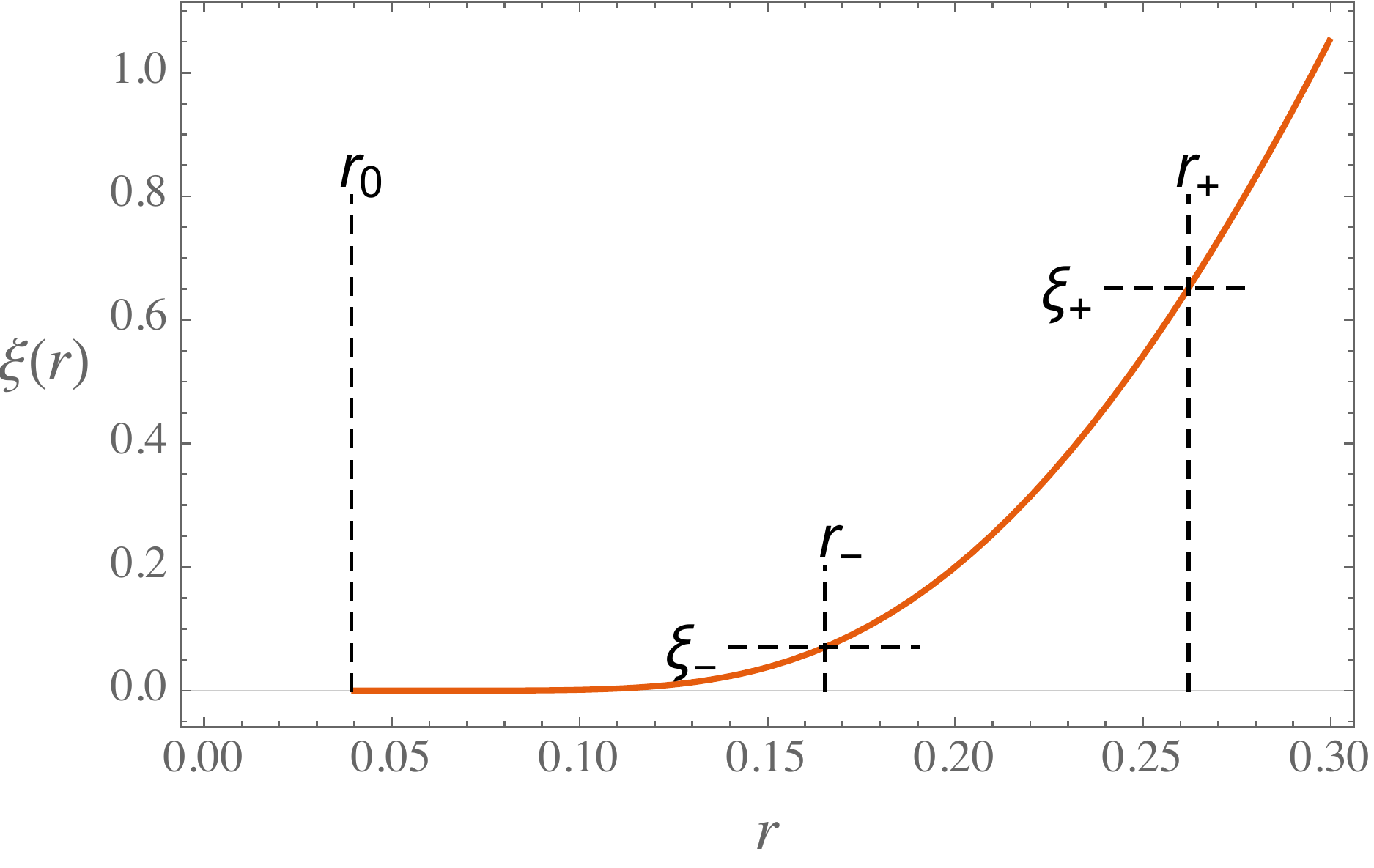}
		\caption{$\xi(r)$}
		\label{fig:Fan-Wang-xi}
	\end{subfigure}
		\begin{subfigure}{.49\textwidth}
	\centering
		\includegraphics[width=\textwidth]{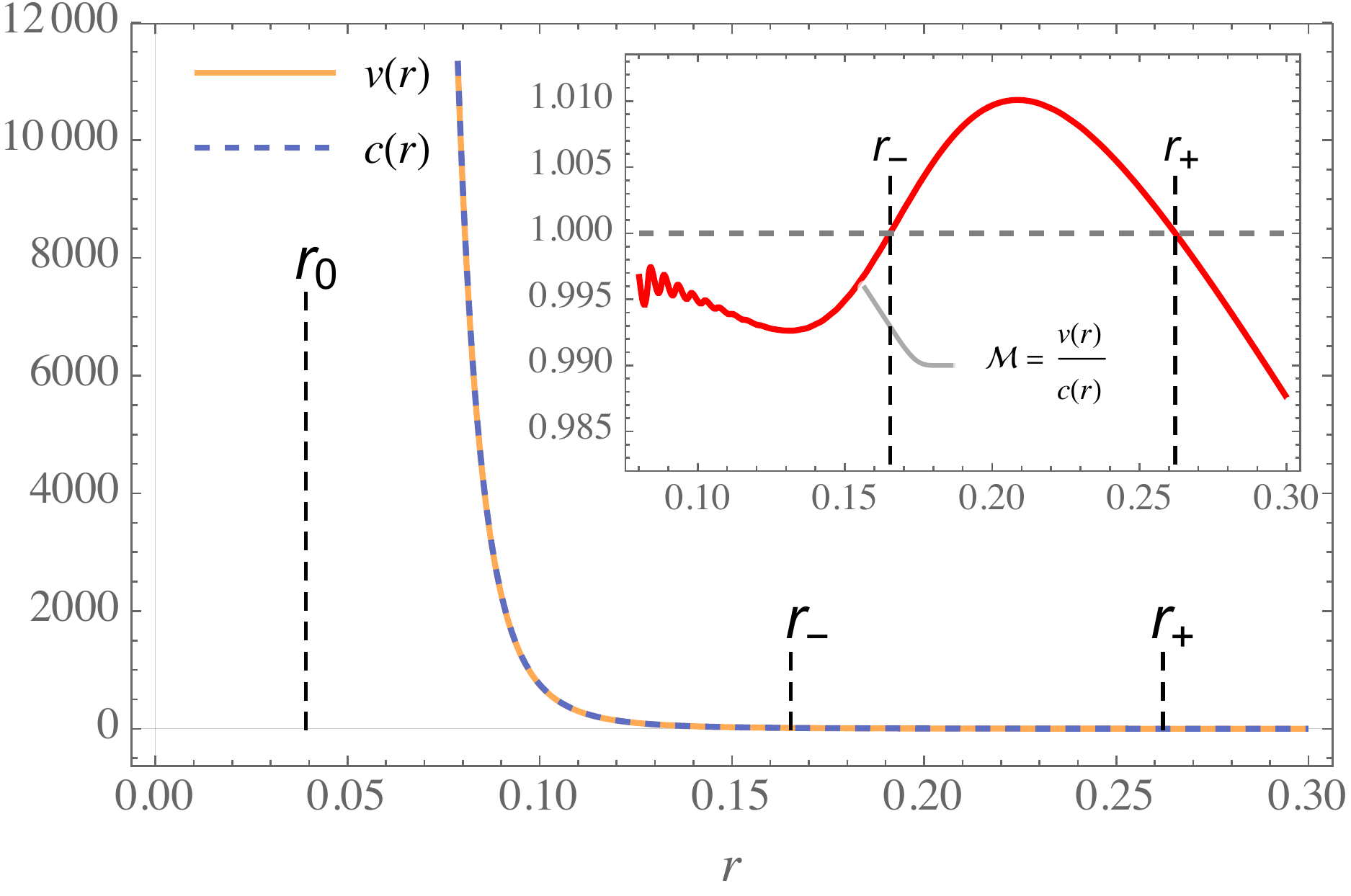}
		\caption{$v(r)$ and $c(r)$}
		\label{fig:Fan-Wang-Mach}
	\end{subfigure}
	\begin{subfigure}{.47\textwidth}
	\centering
		\includegraphics[width=\textwidth]{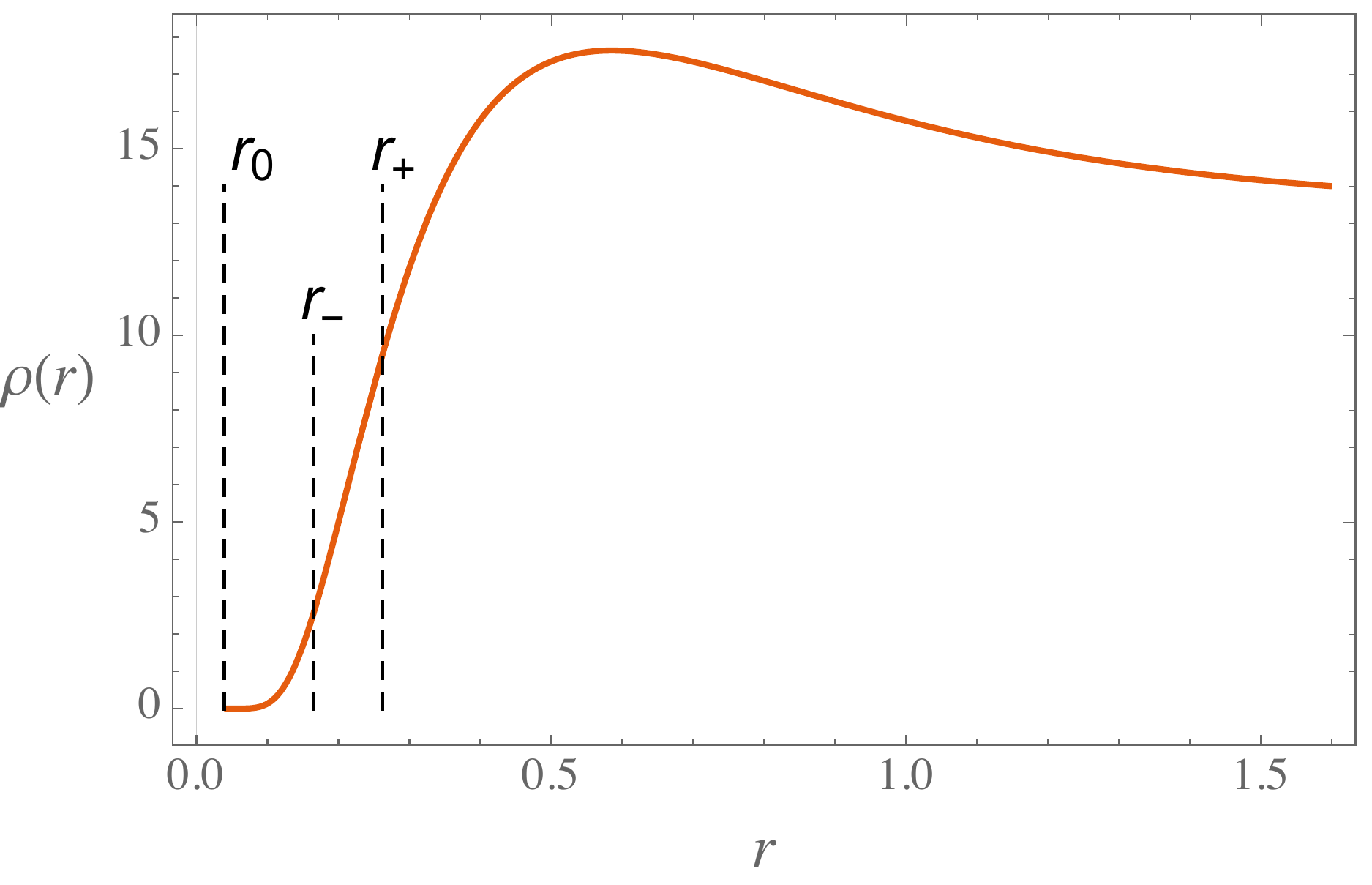}
		\caption{$\rho(r)$}
		\label{fig:Fan-Wang-rho}
	\end{subfigure}
	\begin{subfigure}{.51\textwidth}
	\centering
		\includegraphics[width=\textwidth]{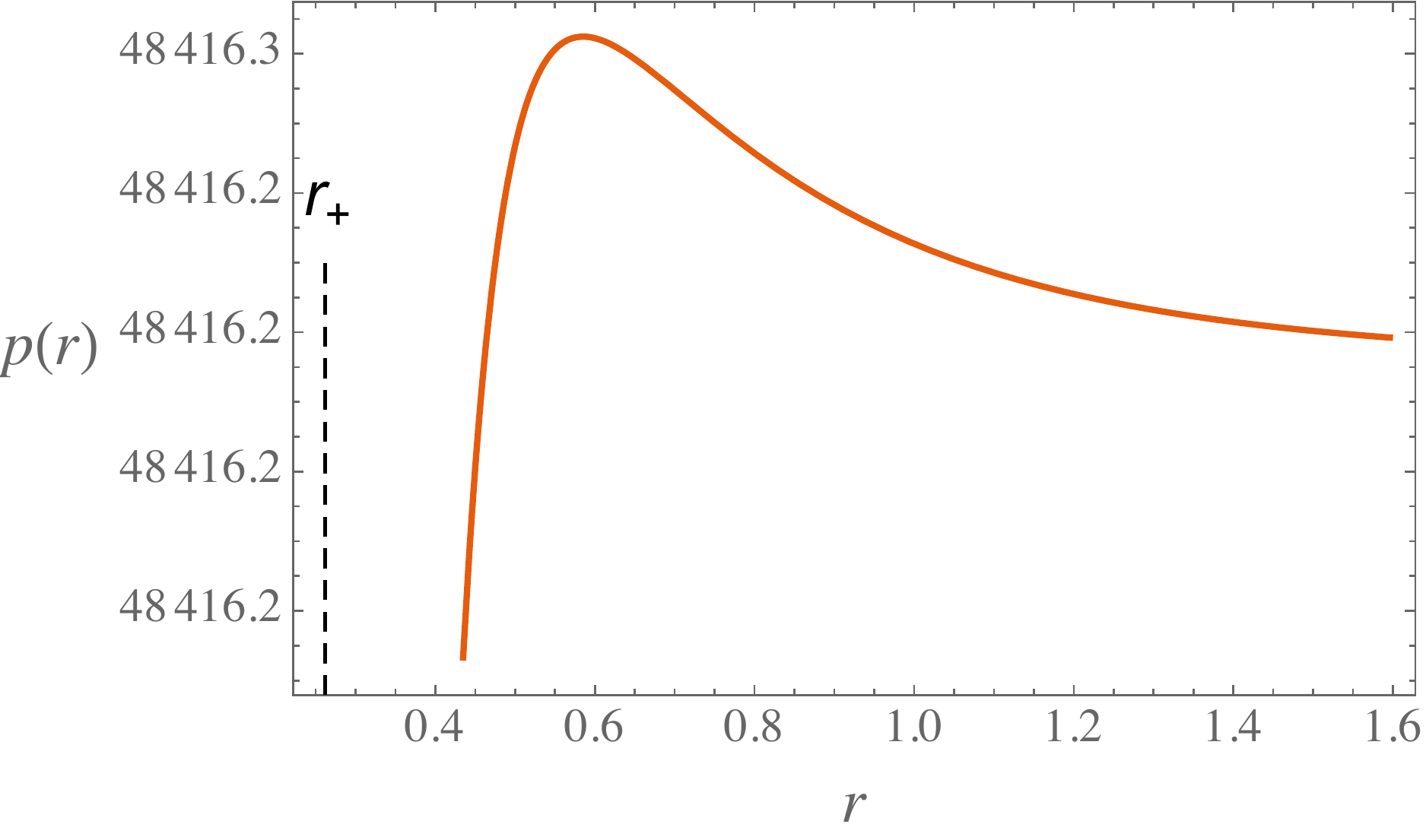}
		\caption{$p(r)$}
		\label{fig:Fan-Wang-p}
	\end{subfigure}
		\begin{subfigure}{.49\textwidth}
	\centering
		\includegraphics[width=\textwidth]{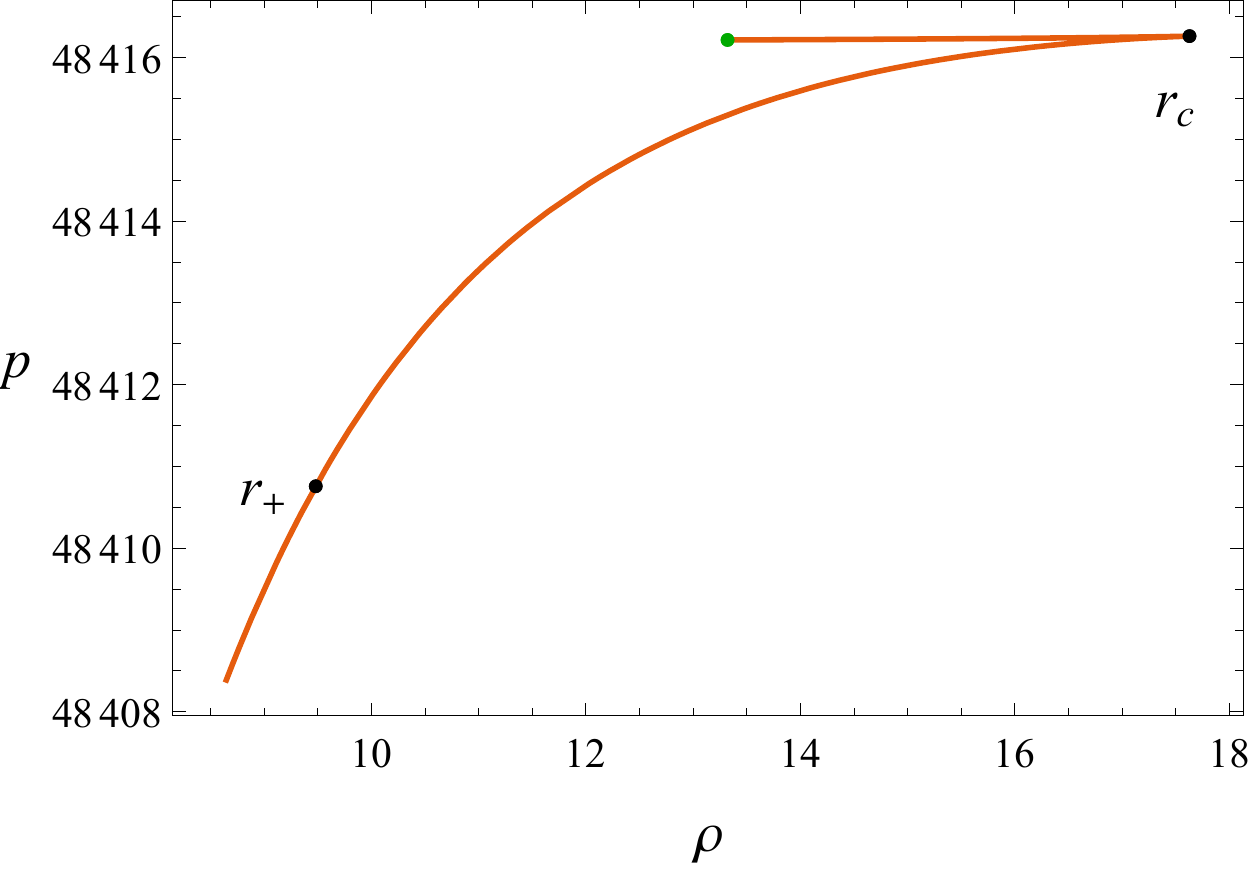}
		\caption{Equation of state}
		\label{fig:Fan-Wang-EoS}
	\end{subfigure}
	\caption{Numerical solutions for the model associated with nonlinear electromagnetic fields, where $q=0.1$, $A=1$,   $\rho_0=1$, and $p_0=0$ are set.}
	\label{fig:fan-wang}
\end{figure}

The numerical results are shown in Fig.\ \ref{fig:fan-wang}, 
where $M=1/2$, $q=0.1$, and $A=1$ are set. 
Because  $p$ and $c$ are divergent at $r=0$, see Eqs.~\eqref{eq:rho-p-fluid} and \eqref{speedlight}.
We have performed a cutoff for the lower boundary by setting $r_0=0.2$ as we did for the remedied Bardeen model.

Here are four points that need to be demonstrated.
\begin{itemize}
	\item 
	The oscillation of Mach numbers at the left tail in Fig.\ \ref{fig:Fan-Wang-Mach} arises from our computational accuracy;
	\item The critical point $r_c$ at which $\rho$ and $p$ are maximized is no longer located between the two horizons (see Figs.\ \ref{fig:Fan-Wang-rho} and \ref{fig:Fan-Wang-p}), which is different from the case of the remedied Bardeen model;
	\item The green dot in Fig.\ \ref{fig:Fan-Wang-EoS} denotes the end of the EoS and corresponds to $\rho\approx 13.322$ and $p\approx 48416.215$;
	\item The upper boundary of Mach numbers can be estimated via a formula similar to Eq.\ \eqref{eq:boundary-mach}, i.e., $\mathcal{M}<1.010$,
	where $z$ reaches its global minimum $z_{\rm min}=-0.010$ at $r=0.209$.
\end{itemize}

%%%%%%%%%%%%%%%%%%%%%%%%%%%%%%%%%%%%%%%%%%%%
\section{Polytropic equations of state}
\label{sec:polytropic}
%%%%%%%%%%%%%%%%%%%%%%%%%%%%%%%%%%%%%%%%%%%%

Motivated by the polytropic behaviors of the equations of state in Sec.\ \ref{sec:simulation},
we try to determine whether the fluid with polytropic equations of state can be used to simulate realistic RBHs.
For this purpose, we suppose that the fluid is barotropic and has  a
polytropic equation of state $p=\tilde B\rho^\gamma$, where $\gamma$ is a real and nonzero number.  Its
valid region is determined below, and $\tilde B$ is constant.
The local speed of sound can be calculated as 
\begin{equation} 
c = B \rho^{(\gamma-1)/2},\qquad B\coloneqq (\tilde B \gamma)^{1/2}>0.\label{speedsound}
\end{equation}
Meanwhile, the fluid should satisfy the continuity equation
which provides the same relationship between $\rho$ and $v$ as Eq.\ \eqref{eq:sol-continue}.
Then, by substituting 
Eqs.~\eqref{speedsound}
and 
 \eqref{eq:sol-continue} into Eq.\ \eqref{eq:metric-original}, we give the metric in terms of $\rho$ as follows:
\begin{equation}
\label{eq:met-poly}
g_{\mu\nu}
=\frac{\rho^{({3-\gamma})/{2}}}{B}
 \left\{-B^2 f \rho^{\gamma -1},\;
 f^{-1},\;
 r^2,\;
 r^2 \sin ^2\theta \right\},\qquad f=1-\frac{A^2} {B^2r^4\rho^{\gamma +1}},
\end{equation}
where $f$ is the shape function.

Now, we investigate whether the above metric can mimic the RHBs with the metric of Eq.\ \eqref{eq:metric-regular}.
To this end, we use the new variable $\xi$ defined by Eq.~(\ref{newvar}), which takes the following form when Eq.~(\ref{speedsound}) is considered: 
\begin{equation}
	\label{newcoord}
	\xi^2=\frac{r^2 \rho^{({3-\gamma})/{2}}}{B}.
\end{equation}
When we replace $r$ with $\xi$ by using Eq.~\eqref{newcoord}, the metric of Eq.\ \eqref{eq:met-poly} becomes
\begin{equation}
g_{\mu\nu}
=
 \diag\left\{-B f \rho^{{(\gamma +1)}/{2}},\frac{f^{-1}\rho^{({3-\gamma})/{2}}}{B (\xi')^2},
 \xi ^2,\xi ^2 \sin ^2\theta\right\},
\end{equation}
where $\xi'\coloneqq \dif \xi/\dif r$. 
Therefore, the condition $g_{tt} g_{\xi\xi}=-1$, leads to the following differential equation of $\rho$:
\begin{equation}
\left[(\gamma -3) r \rho '-4 \rho \right]^2=16 B \rho ^{{(\gamma +5)}/{2}},
\end{equation}
whose general solution is
\begin{equation}
\rho ^{-(\gamma +1)/{4} }=\pm\sqrt{B}+c_8 r^{{(\gamma +1)}/{(3-\gamma)}},
\end{equation}
where $c_8$ is an integration constant. 
Furthermore, we note that the asymptotic behavior of $\rho$ is $\rho\to B^{-2/(\gamma +1)}$ as $r\to0$
if $-1<\gamma <3$,
while for $\gamma <-1\cup\gamma >3$, it is $\rho\to c_8^{-4/(\gamma +1)} r^{-4/(3-\gamma )}$. 
Thus, the asymptotic behaviors of Weyl curvatures for these two cases are  
\begin{subequations}
\begin{equation}
W\sim O(r^{-12})
,\qquad
-1<\gamma <3;
\end{equation}
\begin{equation}
W\sim O\left(r^{-16 (\gamma -1)/(\gamma -3)}\right),\qquad
\gamma <-1\cup \gamma >3.
\end{equation}
\end{subequations}
The asymptotic relation associated with $-1<\gamma<3$ shows that the Weyl curvature inevitably has a singular point at $r=0$;
as for the case of $\gamma <-1\cup \gamma >3$,
the regularity requires $1\leq \gamma <3$, which contradicts  $\gamma <-1\cup \gamma >3$. 
As a result, the fluid with polytropic equations of state cannot simulate the RBHs with the metric of Eq.\ \eqref{eq:metric-regular}.

Let us investigate under what conditions the metric of Eq.\ \eqref{eq:met-poly} 
describes an RBH solution in the whole spacetime.
We calculate the Weyl curvature of the metric of Eq.\ \eqref{eq:met-poly} directly,
\begin{equation}
\begin{split}
\label{eq:poly-weyl}
W =
\frac{\rho^{-\gamma -9}}{12 B^2 r^{12}}
\Big\{
& 2 A^2 (\gamma +5) r^2 \rho '^2
+2 A^2 r \rho  \left[
    (3 \gamma +17) \rho '-2 r \rho ''
    \right]
    +60 A^2 \rho ^2\\
& -B^2 \left(\gamma ^2-4 \gamma +3\right) r^6 \rho ^{\gamma +1} \rho '^2-2 B^2 (\gamma -1) r^5 \rho ^{\gamma +2} \left(r \rho ''-\rho '\right)
\Big\}^2.
\end{split}
\end{equation}
To give the conditions just mentioned, we make an asymptotic ansatz \cite{Frolov:2016pav}  
$\rho\sim r^{-n}$ as $r\to 0$, and substitute it into Eq.\ \eqref{eq:poly-weyl}, where $n$ is a real and positive number.
We find that the  square root of Weyl curvatures consists of the following two terms:
\begin{equation}
r^{-(\gamma -3) n/{2}-2}
\quad \text{and}\quad
r^{(\gamma +5) n/{2} -6},
\end{equation}
where we have omitted the relevant coefficients.
The regularity at $r=0$ provides two inequalities:
\begin{subequations}
\label{eq:rho-div}
\begin{equation}
n\geq \frac{12}{\gamma +5},  \qquad \gamma\in (-5, 1] ;
\end{equation}
\begin{equation}
n\ge -\frac{4}{\gamma -3},  \qquad \gamma\in (1, 3).
\end{equation}
\end{subequations}
On the other hand, the asymptotic flatness requires
\begin{equation} 
\rho\to B^{-{2}/{(\gamma +1)}},
\label{asymflat}
\end{equation}
when $r\to\infty$.
In summary, we conclude that the metric of Eq.\ \eqref{eq:met-poly} describes an RBH 
if Eqs.~(\ref{eq:rho-div}) and (\ref{asymflat}) are satisfied.

We take the Chaplygin gas as an example, where $\gamma=-1$ in Eq.~(\ref{eq:met-poly}) and the density has the form
\begin{equation}
\label{eq:special-solution}
\rho= \rho _0 \frac{l^3 }{r^3}\sqrt{1+\frac{l^2}{r^2}},
\end{equation}
where $l$ is introduced for balancing the length dimension.
The corresponding Weyl curvature is
\begin{equation}
W=\frac{4 r^4 \left[A^2 \left(4 l^2 r^2+l^4\right)+B^2 r^4 \left(8 l^2 r^2+8 l^4+3 r^4\right)\right]^2}{3 B^2 l^{12} \rho _0^4 \left(l^2+r^2\right)^6},
\label{Wyelcuav}
\end{equation}
which has an asymptotic relation 
\begin{equation}
W\sim \frac{4 A^4 r^4}{3 B^2 l^{16} \rho _0^4}+O(r^5),
\end{equation}
when $r\to 0$.
Thus, the Weyl curvature is regular. 
Moreover, the bracket in the denominator of Eq.~(\ref{Wyelcuav}) is an algebraic quadratic function of $r$, but it has no real roots because $l\in \mathbb{R}$ and $B\ne 0$, 
which consequently indicates that the Weyl curvature is finite on the non-negative axis.

It is time for us to investigate the energy conditions for the astronomical counterpart of the metric Eq.\ \eqref{eq:met-poly}.
We study the vacuum equation $T^{t}_{\; t}=T^{r}_{\; r}$ to clarify whether we have to define the energy density and pressure inside and outside the horizon, respectively.
The vacuum equation leads to a second-order nonlinear differential equation of $\rho$:
\begin{equation}
4 \rho  \left[2 (\gamma -1) \rho '+(\gamma -3) r \rho ''\right]-r\,(\gamma -3) (\gamma +5)(\rho ')^2=0,
\end{equation}
whose general solution is
\begin{equation}
\rho = c_{10} r^{{(\gamma -3)}/{4}}
\left(3-\gamma -4 c_9 r^{{(\gamma +1)}/{(\gamma -3)}}\right)^{-{4}/{(\gamma +1)}},
\end{equation}
where $c_9$ and $c_{10}$ are integration constants. 
Furthermore, the asymptotic analysis at $r\to 0$ gives us two situations:
\begin{subequations}
\begin{equation}
\rho\sim O(1),\quad
W\sim O( r^{-12}),\quad \text{for }\;  -1<\gamma <3;
\end{equation}
\begin{equation}
\rho \sim  O(r^{4/(\gamma -3)}),\quad
W\sim O(r^{-16 (\gamma -1)/(\gamma -3)}),\quad 
\text{for }\; \gamma <-1.
\end{equation}
\end{subequations}
None of them can realize a regular Weyl curvature at $r=0$; i.e.,
we have to discuss the energy conditions inside and outside a horizon separately.

The definitions of the energy density $\epsilon$ and radial pressure $p_r$ depend on the number of horizons $\mathfrak{n}$;
meanwhile, the horizons separate the spacetime into $\mathfrak{n}+1$  regions.
If we start from the region outside the outermost horizon and denote that area as $1$,
then for the region with odd number $\mathfrak{n}\in2\mathbb{N}+1$, 
the energy density and radial pressure 
are defined by
\begin{equation}
\epsilon^{\rm odd}=-\frac{G^{t}_{\;t}}{8\pp},\qquad
p_r^{\rm odd} = \frac{G^{r}_{\;r}}{8\pp};
\end{equation}
while for even $\mathfrak{n}\in2\mathbb{N}$, they are defined by
\begin{equation}
\epsilon^{\rm even}=-\frac{G^{r}_{\;r}}{8\pp},\qquad
p_r^{\rm even} = \frac{G^{t}_{\;t}}{8\pp}.
\end{equation}
It can be verified that the model of Eq.\ \eqref{eq:special-solution} violates the DEC because there is no intersection between $\epsilon\ge \abs{p_r}$ and $\epsilon\ge \abs{p_t}$,
regardless of whether the number of horizons is odd or even. Thus, the Chaplygin gas cannot be used to mimic an astronomical counterpart.
In fact, the inverse problem, i.e., constructing $\rho$ from the energy conditions, 
is rather complicated because the DEC leads to four second-order nonlinear differential inequalities,
which are difficult to deal with. 
Therefore, we stop searching for the models with a polytropic EoS and leave this for future studies.

%%%%%%%%%%%%%%%%%%%%%%%%%%%%%%%%%%%%%%%%%%%%%%%%%%%%%%%%%%%%%%%%%%%%%%%%
\section{Cylindrical regular black holes and their equatorial sections}
\label{sec:cylinder}
%%%%%%%%%%%%%%%%%%%%%%%%%%%%%%%%%%%%%%%%%%%%%%%%%%%%%%%%%%%%%%%%%%%%%%%%

In this section, we study the RBHs with cylindrical symmetry.
The metric can be cast as follows:
\begin{equation}
\label{eq:black-rings}
\dif s^2 = 
	-f\dif t^2+f^{-1} \dif \xi^2
	+\xi^2 \dif \phi^2 +\me^{2\zeta(\xi)}\dif z^2,\qquad f=1-2M \sigma/\xi,
\end{equation}
where $\zeta$ is a real function of $\xi$.
Instead of analyzing the Weyl curvature, we analyze the Ricci scalar because of its simplicity:
\begin{equation}
\label{eq:ricci-cylinder}
R=
\frac{4 M \zeta'\sigma'}{\xi}+\frac{4 M \sigma \zeta'^2}{\xi}+\frac{4 M \sigma \zeta''}{\xi}+\frac{2 M \sigma''}{\xi}-2 \zeta'^2-\frac{2 \zeta '}{\xi}-2 \zeta '',
\end{equation}
where the prime denotes the derivative with respect to $\xi$. 
The Ricci scalar is regular at $\xi=0$ if $\sigma$ and $\zeta$ have the asymptotic forms
$\sigma\sim O(\xi^m)$ and $\zeta\sim O(\xi^n)$, where
$m\ge 3$ and $n\ge 2$ or $n=0$.
In contrast, the asymptotic flatness demands 
$\sigma\sim O(\xi^{\tilde m})$ with $\tilde m<1$ and $\zeta\sim O(1)$.

Now, let us consider the energy conditions of this type of black hole.
When $\zeta=0$, $G^t_{\; t}=G^\xi_{\; \xi}$ is valid in the regions inside and outside a horizon.
The energy density and pressures can be calculated only in terms of $\sigma$ and its derivatives:
\begin{equation}
\begin{split}
\epsilon=\frac{M }{8 \pp  \xi^3}\left(\xi \sigma '-\sigma \right),\qquad
p_\xi = \frac{M }{8 \pp  \xi^3}\left(\sigma-\xi \sigma '\right),\\
p_\phi = -\frac{M}{8 \pp  \xi^3} \left[\xi \left(\xi \sigma ''-2 \sigma '\right)+2 \sigma \right],\qquad
p_z =-\frac{M \sigma ''}{8 \pp  \xi}.
\end{split}
\end{equation}
Thus, the DEC is 
\begin{equation}
\xi \left(\xi \sigma ''-\sigma '\right)= \sigma,\qquad  \xi \left(\xi \sigma ''+\sigma '\right)\ge \sigma. 
\end{equation}
Note that the equality comes from $\epsilon\ge p_\phi \cap p_z\ge -\epsilon$ 
and can be used to fix $\sigma$, i.e.,
$\sigma = c_{11} \xi+c_{12} \xi \ln (\xi)$, where $c_{11}$ and $c_{12}$ are integration constants. 
However, the corresponding Ricci scalar, i.e., $R=2 c_{12}M/\xi^2$, is singular unless the metric is trivial, $c_{12}=0$.
In other words, the metric Eq.\ \eqref{eq:black-rings} with $\zeta=0$ can never represent a realistic RBH.

By omitting the flow along the $z$ direction, i.e., considering only the equatorial section (ES), 
we reduce the DEC to 
\begin{equation}
\epsilon\ge p_\xi \ge -\epsilon,\qquad
\epsilon\ge p_\phi \ge -\epsilon,\qquad
\epsilon\ge 0,
\end{equation}
where the contribution from $p_z$ is ignored.
Therefore, we have the following three differential inequalities:
\begin{equation}
\xi^2 \sigma ''+\sigma\ge \xi \sigma ',\qquad
 \xi^2 \sigma ''+3 \sigma \le 3 \xi \sigma ',\qquad
 \xi \sigma '\ge\sigma.\label{cylininequalES}
\end{equation}
We take the modified Hayward BH as an example (see Eq.~\eqref{modhayward}), but discuss its cylindrical counterpart:
\begin{equation}
\sigma=\frac{M^{\alpha-3}\xi^3}{q^{\alpha }+\xi^{\alpha }}.\label{cylinsigma}
\end{equation}
By substituting Eq.~(\ref{cylinsigma}) into Eq.~(\ref{cylininequalES}), we obtain
\begin{equation}
\begin{split}
2 q^{\alpha }\geq (\alpha -2) \xi^{\alpha },
\qquad
(\alpha +2) q^{\alpha }\ge (\alpha -2) \xi^{\alpha },\\
\left[2 q^{\alpha }+(\alpha -2) \xi^{\alpha }\right]^2\ge [\alpha  (\alpha +8)-16] q^{\alpha } \xi^{\alpha },
\end{split}
\end{equation}
which hold for $\xi\in [0,\infty)$ if $-4 (\sqrt{2}+1)\le \alpha \leq 4 (\sqrt{2}-1)$.

\begin{figure}
\centering
	\begin{subfigure}{.51\textwidth}
	\centering
		\includegraphics[width=\textwidth]{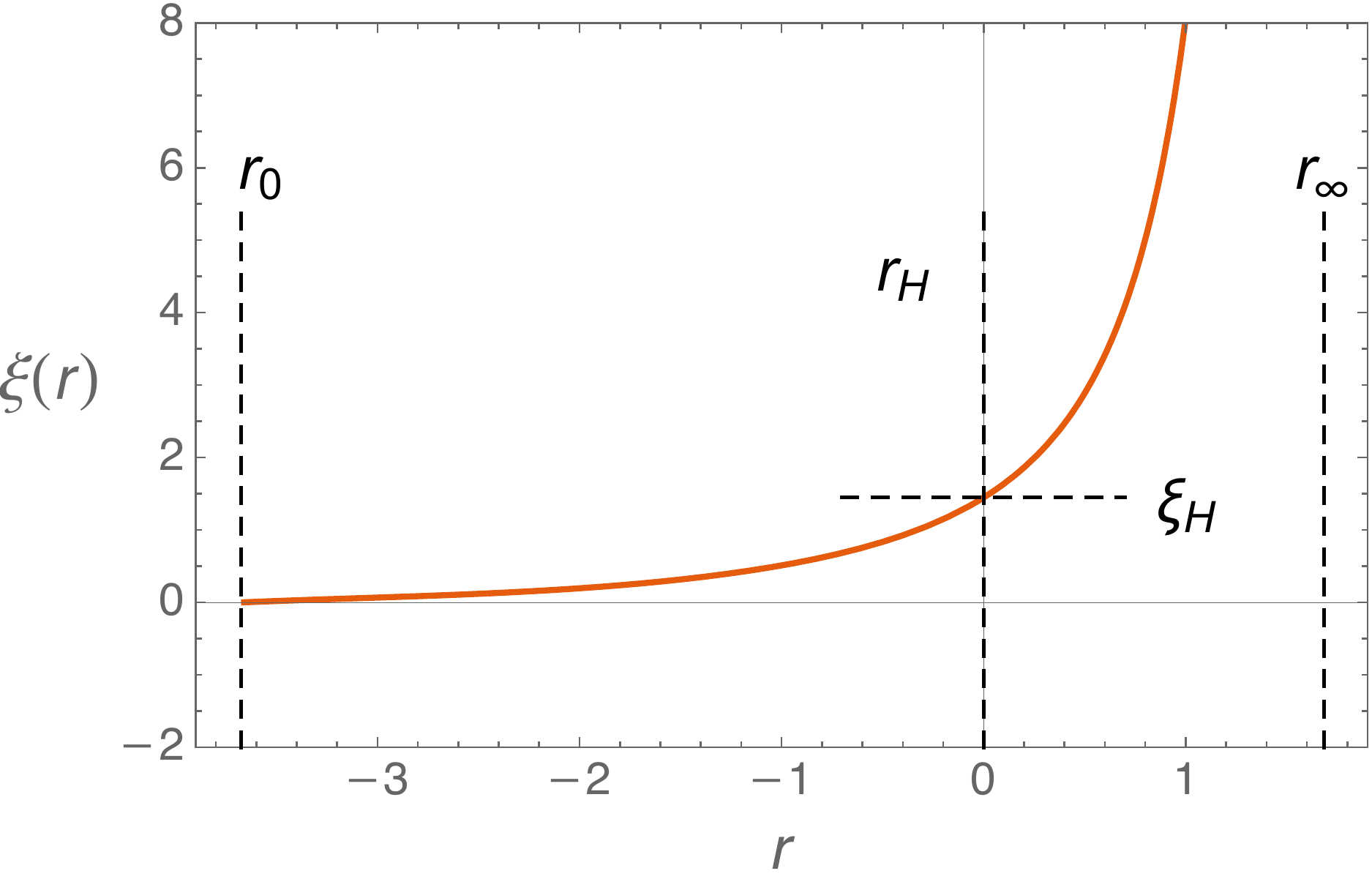}
		\caption{$\xi(r)$}
		\label{fig:es-bardeen-xi}
	\end{subfigure}
		\begin{subfigure}{.47\textwidth}
	\centering
		\includegraphics[width=\textwidth]{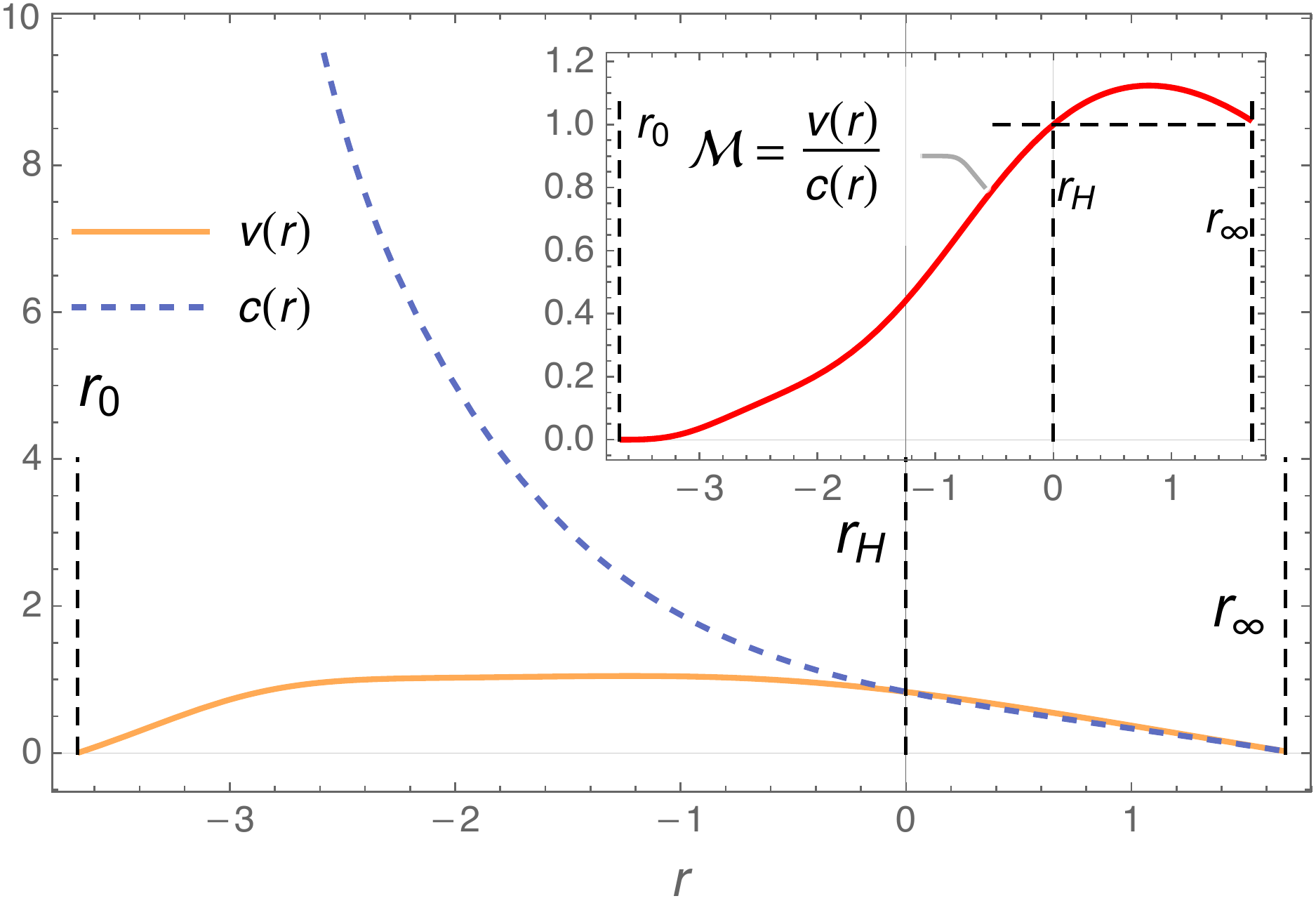}
		\caption{$v(r)$ and $c(r)$}
		\label{fig:es-bardeen-Mach}
	\end{subfigure}
	\begin{subfigure}{.49\textwidth}
	\centering
		\includegraphics[width=\textwidth]{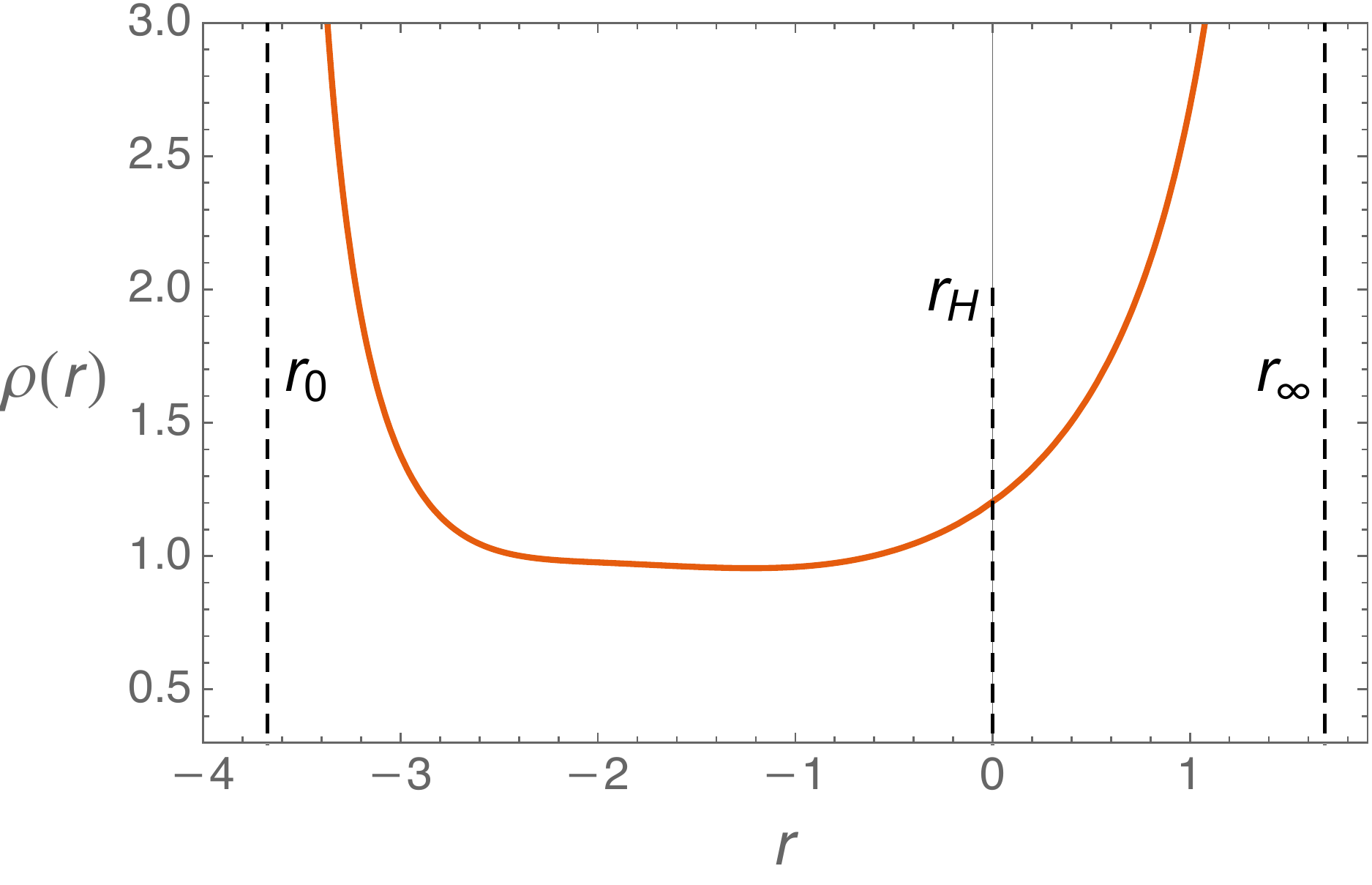}
		\caption{$\rho(r)$}
		\label{fig:es-bardeen-rho}
	\end{subfigure}
\begin{subfigure}{.49\textwidth}
	\centering
		\includegraphics[width=\textwidth]{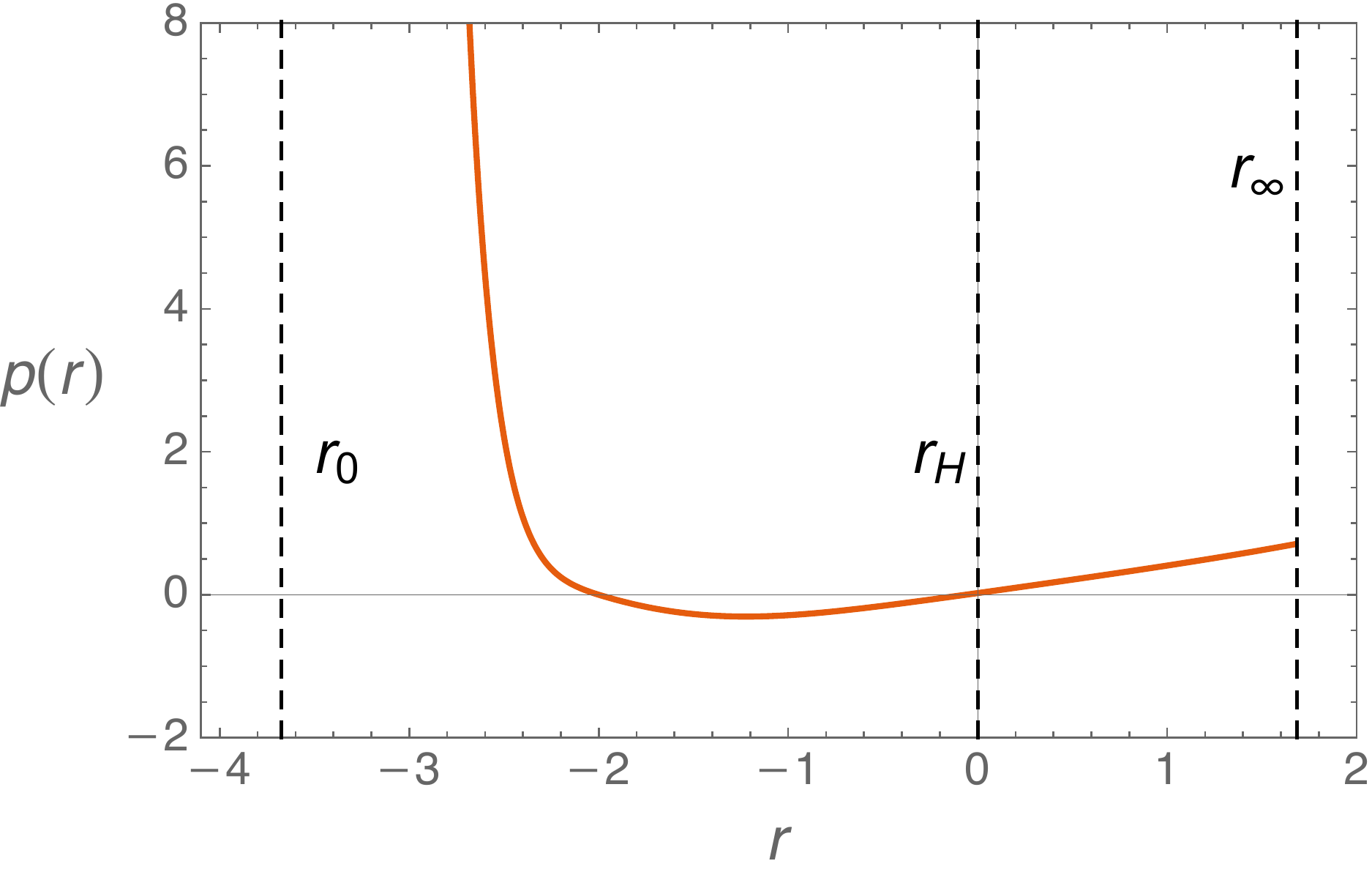}
		\caption{$p(r)$}
		\label{fig:es-bardeen-p}
	\end{subfigure}
	\begin{subfigure}{.49\textwidth}
	\centering
		\includegraphics[width=\textwidth]{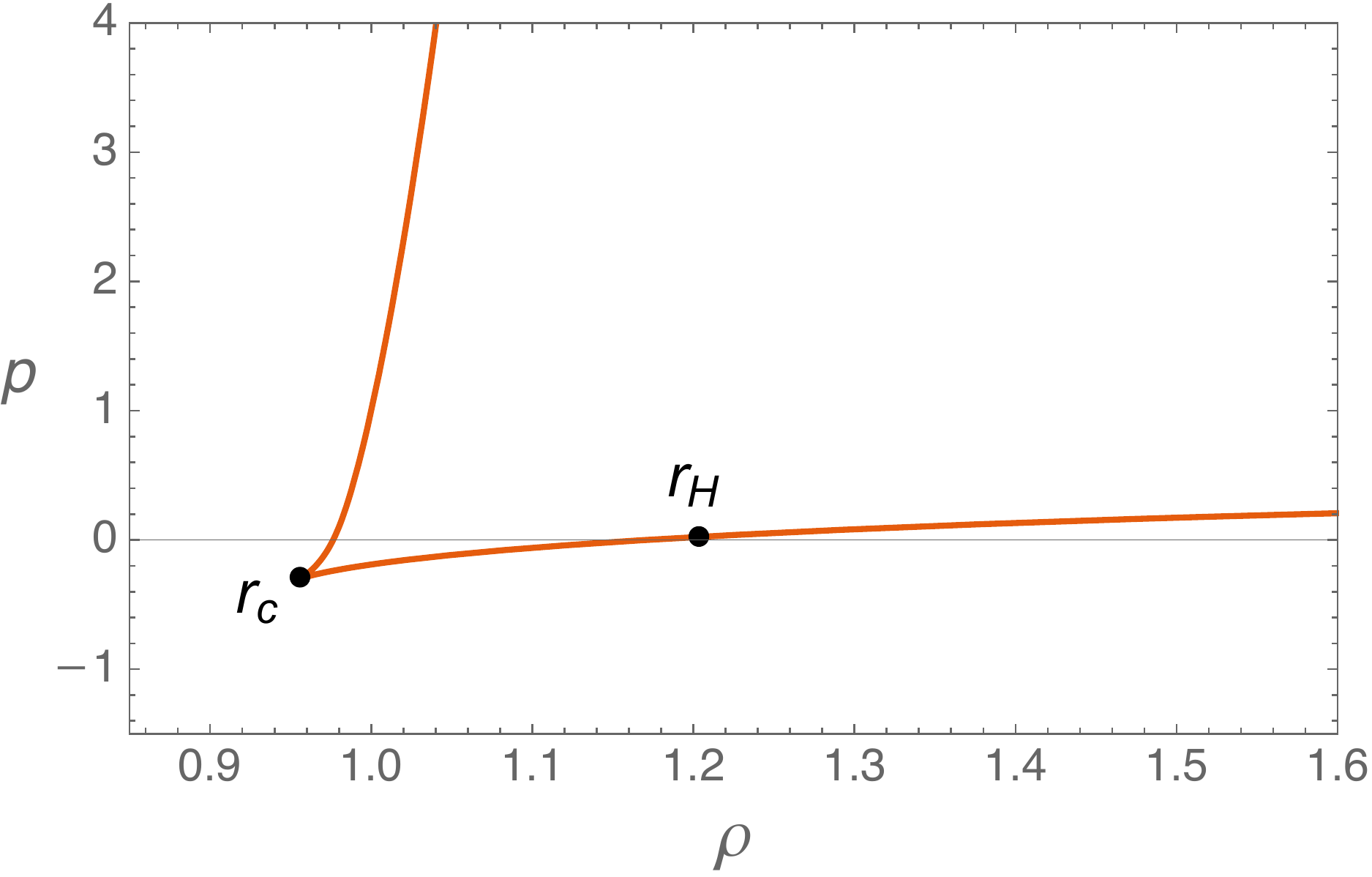}
		\caption{Equation of state}
		\label{fig:es-bardeen-EoS}
	\end{subfigure}
	\caption{Numerical solutions for the ES of repaired cylindrical Hayward-like models with $M=q=0.5$ and $A=1$, where $r_c\approx -1.352$ and 
	$r_0\approx-3.675$, which is determined by $\xi(r_0)=0$, and $r_\infty\approx 1.684$, which is determined by a very large value of $\xi$ in the numerical calculation.}
	\label{fig:es-bardeen}
\end{figure}

To simulate the equatorial sections of cylindrical regular black holes, we take
$\alpha = 3/2$ in Eq.~\eqref{cylinsigma}. Then, we can obtain the nonlinear differential equation (see Eq.~(12) of Ref.\ \cite{deOliveira:2021edr})
that describes the relationship between the radial coordinate  of black holes ($\xi$) and that of the simulation in fluids ($r$): 
\begin{equation}
(\xi ')^4-f(\xi) \xi ^2 (\xi ')^2-A^2\xi^6=0,\qquad 
f(\xi)=1-\frac{2 M^{-1/2} \xi^2}{q^{3/2}+\xi^{3/2}},\label{nonlinearEQ}
\end{equation}
where the prime denotes the derivative with respect to $r$. 
From Eq.~(\ref{nonlinearEQ}), we can obtain the asymptotic relations of $\xi$ and $r$,
\begin{equation}
\label{eq:es-zero}
\xi^\pm\sim c_{13} \me^{\pm r}, \qquad
\text{as}\quad \xi\to0,
\end{equation}
and
\begin{equation}
\label{eq:es-inf}
\xi^\pm \sim\frac{4}{(\sqrt{A} r\pm c_{14})^2}, \qquad
\text{as}\quad \xi\to\infty,
\end{equation}
where $c_{13}$ and $c_{14}$ are integration constants. 
Meanwhile, because the physical variables of fluids can be represented by $\xi$
and its derivatives, see Eq.~(\ref{eq:phys-var}), 
\begin{equation}
\label{eq:phy-var-es}
c= \frac{\xi '}{\xi ^2},\qquad
v=\frac{A\xi }{\xi '},\qquad \rho= \frac{\xi '}{\xi},
\qquad
p'=\frac{(\xi ')^2 }{\xi^6}\left[\xi \xi ''-(\xi ')^2\right],
\end{equation}
their asymptotic behaviors for the case of the positive sign are 
\begin{equation}
c^+\to \frac{\me^{-r}}{c_{13}},\qquad
v^+\to A,\qquad \rho^+\to 1,\qquad
p'^+\to 0,
\end{equation}
when $r\to 0$ and 
\begin{equation}
\begin{split}
c^+\to -\frac{\sqrt{A}}{2}  \left(\sqrt{A} r+c_{14}\right),&\qquad
v^+\to -\frac{\sqrt{A}}{2}  \left(\sqrt{A} r+c_{14}\right),\\
\rho ^+\to -\frac{2 \sqrt{A}}{\sqrt{A} r+c_{14}},&\qquad 
p'^+\to \frac{A^2}{2},
\end{split}
\end{equation}
when $r\to \infty$.
The phenomenon of transonic flows occurs outside the horizon, and
the Mach number converges to $1$, i.e., $\mathcal{M}\to1$, as $r$ approaches infinity.
The corresponding numerical analysis is shown in Fig.\ \ref{fig:es-bardeen}, 
where we have adopted the setting $M=1/2$, $q=1/2$, and $A=1$ 
and chosen $r_{\rm H}=0$, which corresponds to  $\xi(0)\approx -3.837$. 
The upper boundary of the Mach numbers is determined numerically, i.e., $\mathcal{M}\le 1.124$,
and the maximum is reached at $r\approx0.796$.

For the case of $\zeta\ne 0$, the vacuum equation $G^t_{\;t}=G^\xi_{\; \xi}$ gives rise to  
\begin{equation}
\zeta '^2+\zeta ''=0,
\end{equation}
which is valid for an arbitrary $\xi\in [0,\infty)$.
The solution $\zeta =\ln (\xi-c_{15})+c_{16}$, where $c_{15}$ and $c_{16}$ are integration constants, does not satisfy the regular condition.
Thus, we have to separate the discussion for the inside of the horizon from that for the outside of the horizon.
However, $p_\xi^{\rm out}\ge -\epsilon^{\rm out}$ and $p_\xi^{\rm in}\ge -\epsilon^{\rm in}$
provide the same inequality
\begin{equation}
\label{eq:inq-zeta}
\zeta '^2+\zeta ''\le 0,
\end{equation}
which can be solved by multiplying both sides by $\exp(\zeta)$, 
\begin{equation}
\frac{\dif^2 }{\dif \xi^2} \exp(\zeta)\le 0.
\end{equation}
This inequality indicates that $\exp(\zeta)$ is a concave function in 
$\xi\in [0, \xi_{\rm H})\cup (\xi_{\rm H}, \infty)$; thus, $\zeta$ is a concave function,
because the logarithm of a non-negative and concave function is concave.\footnote{Suppose $y(x)\ge0$ and $y''(x)<0$, thus $\dif^2\ln(y)/\dif x^2=(-y'^2+y y'')/y^2$ is negative, i.e., $\ln(y)$ is concave.}
This can also be seen from Eq.\ \eqref{eq:inq-zeta}; $\zeta''$ is nonpositive because $(\zeta')^2$ is nonnegative. 
Nevertheless, $\zeta$ is bounded to satisfy the condition of finite curvatures, 
which contradicts the concavity of $\zeta$. 
In other words,  the metric Eq.\ \eqref{eq:black-rings} can in no way represent a realistic RBH.
In contrast, the $2$D BHs with polar symmetry can be regarded as cylindrical BHs with the $z$ direction being suppressed.
In the following section, we investigate the $2$D polar-symmetric RBHs and their simulations in fluids.

%%%%%%%%%%%%%%%%%%%%%%%%%%%%%%%%%%%%%%%%%%%%%%%%%%%%%%%%%%%%%%%%%%%%%%%%%%%
\section{Simulations of lower dimensional regular black holes}
\label{sec:low-dimension}
%%%%%%%%%%%%%%%%%%%%%%%%%%%%%%%%%%%%%%%%%%%%%%%%%%%%%%%%%%%%%%%%%%%%%%%%%%%

The $(2+1)$D simulation is rather different from the $(3+1)$D case discussed in the previous section.
First,  the Weyl curvature tensor vanishes identically for any $(2+1)$D spacetime, 
thus the Weyl scalar is no longer an appropriate candidate for analyzing the curvature divergence.
Second, the regularity condition is closely related to the dimension of spacetime.
When one studies the RBHs in a $(2+1)$D spacetime, the criteria for shape functions will change, see App.\ \ref{app:high-rbh}.
Thirdly, the form of acoustic line elements also depends on dimension, in particular the prefactor \cite{Barcelo:2005fc}.
 Moreover, the $(2+1)$D analogue BHs in a fluid are rather interesting and relatively easy to be realized in a laboratory.

Let us start considering a circularly symmetric BH, 
\begin{equation}
\bar g_{ij}=\diag\{-f, f^{-1}, \xi^2\},\qquad
f=1-\mu \sigma(\xi),\label{2dbh}
\end{equation} 
where $\mu$ is mass-like parameter.
According to App.\ \ref{app:high-rbh}, such a metric is curvature regular at the BH center
if $\sigma\sim O(\xi^n)$, $n\ge 2$, as $\xi\to 0$, 
and asymptotic flat if $\sigma \to 0$ as $\xi\to \infty$.
In order to construct an example that satisfies these conditions, we make an ansatz,
\begin{equation}
\label{eq:ansatz-low}
\sigma(\xi) = \frac{\xi^\alpha}{\xi^\beta+q^\beta}\quad \text{with}\quad
\alpha,\;\beta\ge 0,
\end{equation}
and substitute it into Kretschmann scalars. 
We find that the Kretschmann scalar is regular at the BH center if $\alpha\ge 2$ and that 
 the asymptotic flatness is satisfied if $\beta>\alpha$.
Next, given a particular $\alpha$ we are going to fix  $\beta$ by the DEC.
For instance, taking $\alpha=2$, we reduce the DEC to the two inequalities,
\begin{equation}
(\beta +2) q^{\beta }\ge (\beta -2) \xi^{\beta },\qquad
4 q^{2 \beta }+(\beta -2)^2 \xi^{2 \beta }\ge [\beta  (\beta +4)-8] q^{\beta} \xi^{\beta }.
\end{equation}
Because the left hand sides of the two inequalities are positive, the two inequalities hold for all non-negative $\xi$ and $q$ 
if their right hand sides are non-positive, 
i.e., the DEC is satisfied in the whole spacetime.
Following this idea, we find $0<\beta \le 2 \sqrt{3}-2\approx 1.46$. 
However, this result contradicts to the asymptotic flatness.
In fact, for the case of $\alpha\ge 2$, the positive energy density $\epsilon \ge 0$ leads to  $\alpha\ge \beta$ in $\xi\in[0,\infty)$,
while the asymptotic flatness requires $\beta>\alpha$. No intersections exist.

If relaxing the asymptotic flatness, we replace it with the Ricci flatness, $R=0$, at infinity.
Then substituting the ansatz Eq.\ \eqref{eq:ansatz-low} into the Ricci scalar, we obtain
\begin{equation}
\frac{R}{\mu} =
\frac{ 
(\beta -3) (\beta -2) \xi^{2 \beta }
+[12-\beta  (\beta +5)] q^{\beta } \xi^{\beta }
+6 q^{2 \beta }}
{\left(\xi^{\beta }+q^{\beta }\right)^3}.
\end{equation}
Since $\beta$ is non-negative, 
the power of $\xi$ in the denominator is larger than that in the numerator,
thus $R$ vanishes as $\xi$ approaches infinity.
In other words, if $\alpha=2$, the metric with Eq.\ \eqref{eq:ansatz-low} automatically satisfies the condition of Ricci flatness. As a result, the model Eq.\ \eqref{eq:ansatz-low} together with the Ricci flatness is regular and satisfies the DEC, that is, it is a realistic RBH.

Furthermore, before we focus on the analogue in a fluid, 
we make a note on the toy model we just constructed, see Eqs.~(\ref{2dbh}) and (\ref{eq:ansatz-low}).
The causal structure of the $(2+1)$D RBH with Eq.\ \eqref{eq:ansatz-low} is exotic. 
Since the power of $\xi$ in the numerator of $\sigma$ is larger than that of the denominator, i.e., $\alpha=2$ and $0< \beta \le 2 \sqrt{3}-2\approx 1.46$, 
$\sigma$ is an increasing function with respect to $\xi$. 
Thus, the shape function $f$ is greater than zero inside the horizon but less than zero outside the horizon.
This indicates that this $(2+1)$D RBH has an opposite structure of lightcones when compared with that of usual BHs, like the Schwarzschild BH.

Now let us turn to the simulation. From Eq.~(\ref{eq:master}), we obtain~\cite{deOliveira:2021edr} the relation between $\xi$ and $r$, 
\begin{equation}
(\xi')^4-f(\xi) \xi^2 (\xi')^2-A^2 \xi^6=0.
\end{equation}
For a specific case, $\alpha=2$ and $\beta=1$, i.e., $\sigma= \xi^2/(\xi+q)$, we find the asymptotic solutions, 
\begin{equation}
\label{eq:2d-zero}
\xi^\pm_0(r)\sim c_{17} \me^{\pm r}, \quad
\text{as}\quad \xi\to0,
\end{equation}
and
\begin{equation}
\label{eq:2d-inf}
\xi^\pm_\infty(r) \sim\frac{4}{(\mp k r+c_{18})^2}, \quad
\text{as}\quad \xi\to\infty,
\end{equation}
where $k:=\sqrt{(\sqrt{4 A^2+\mu ^2}-\mu )/2}$, and $c_{17}$ and $c_{18}$ are integration constants.
Since the situation with the positive sign corresponds to the positive correlation between $\xi$ and $r$, 
we would like to select it as the candidate for the simulation. 
Meanwhile, we note from Eq.\ \eqref{eq:2d-inf} that 
there is a movable singularity in the asymptotic solution as $\xi\
\to \infty$. 
In other words, $\xi^+_\infty(r)$ diverges at a finite $r_\infty$, where $r_\infty$  depends on the choice of $\xi^+_\infty(0)$.
To estimate the value of $r_\infty$, 
we apply the condition $\xi^+_\infty(0)=\xi_{\rm H}$
which determines the integration constant $c_{18}=2 \sqrt{\sqrt{2}-1}$ and $r_\infty = 2 \sqrt{(\sqrt{2}-1)/{(\sqrt{4 A^2+\mu ^2}-\mu )}}$. Here $\xi_{\rm H}$ denotes the horizon radius of the astronomical counterpart depicted by Eq.~(\ref{2dbh}).

\begin{figure}[!ht]
\centering
	\begin{subfigure}{.51\textwidth}
	\centering
		\includegraphics[width=\textwidth]{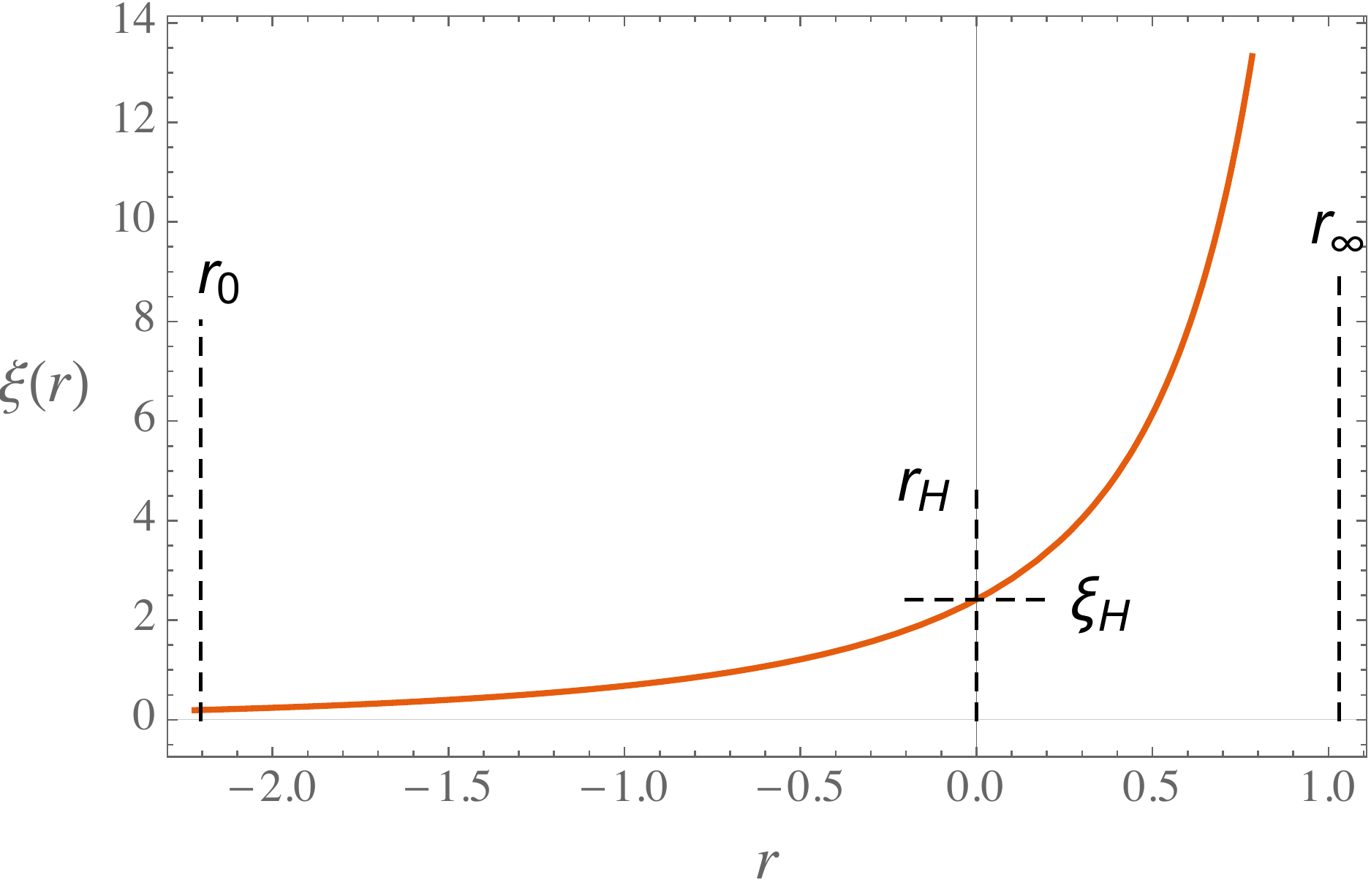}
		\caption{$\xi(r)$}
		\label{fig:2d-bardeen-xi}
	\end{subfigure}
		\begin{subfigure}{.47\textwidth}
	\centering
		\includegraphics[width=\textwidth]{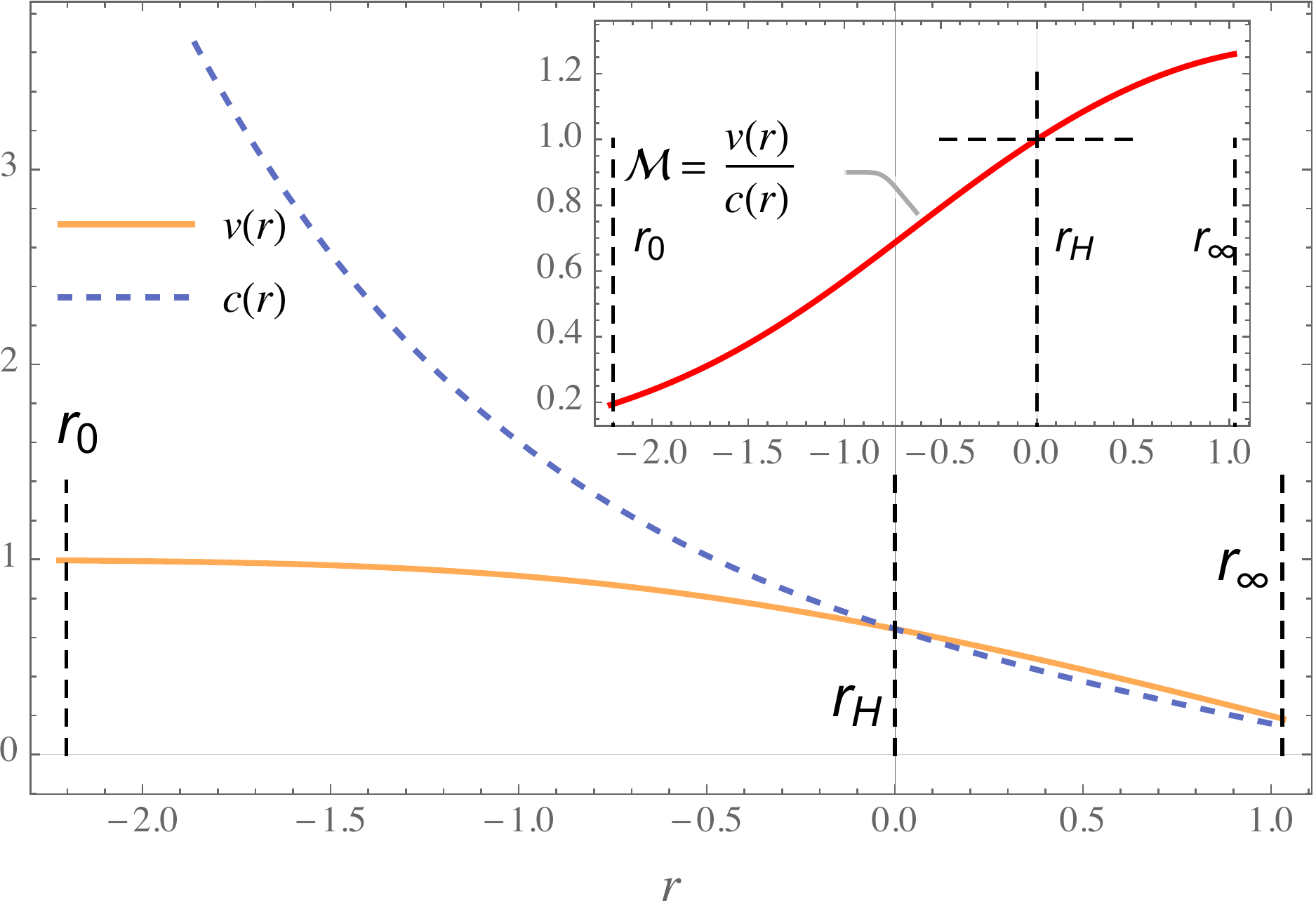}
		\caption{$c(r)$ and $v(r)$}
		\label{fig:2d-bardeen-Mach}
	\end{subfigure}
	\begin{subfigure}{.49\textwidth}
	\centering
		\includegraphics[width=\textwidth]{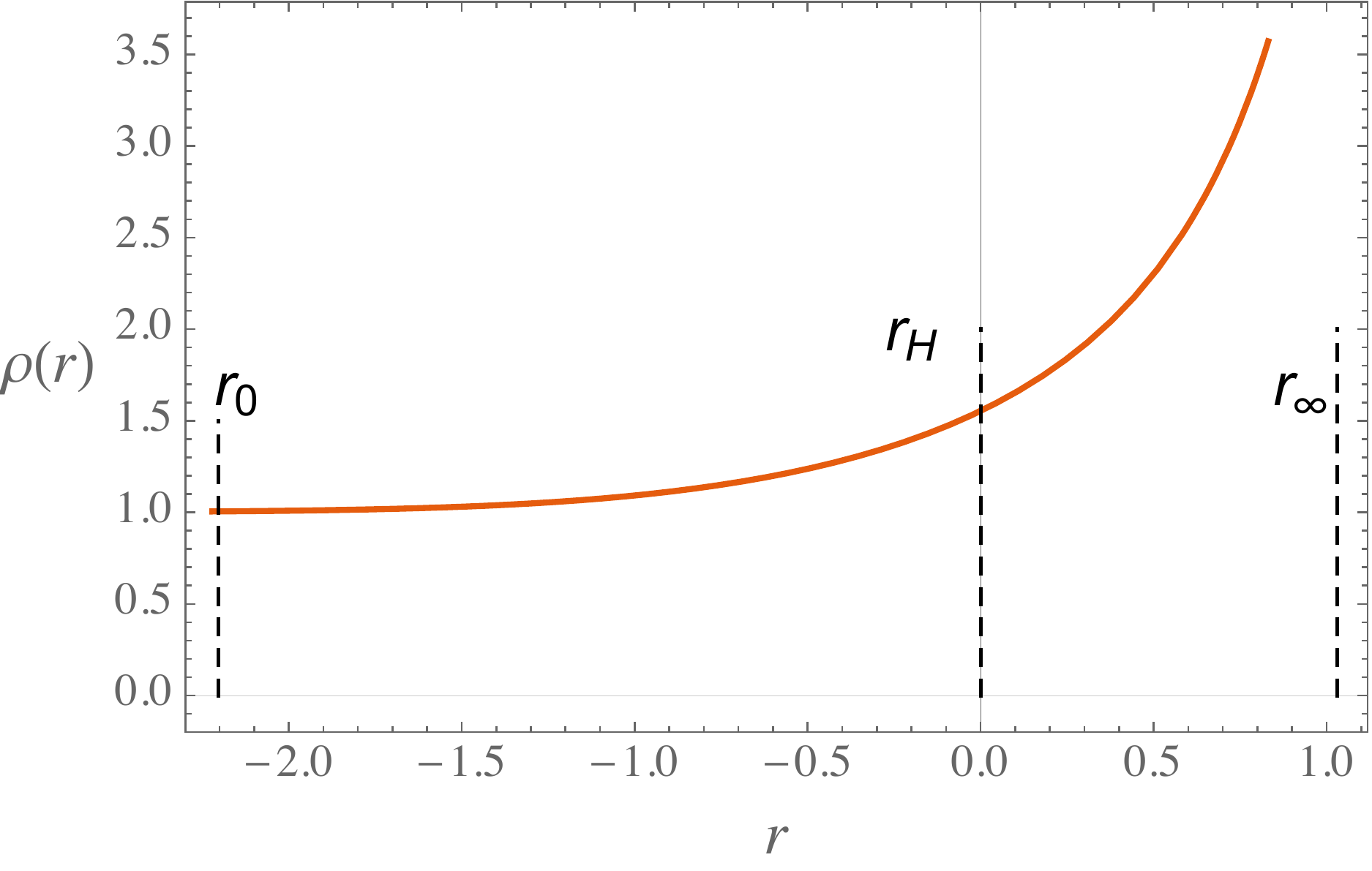}
		\caption{$\rho(r)$}
		\label{fig:2d-bardeen-rho}
	\end{subfigure}
\begin{subfigure}{.49\textwidth}
	\centering
		\includegraphics[width=\textwidth]{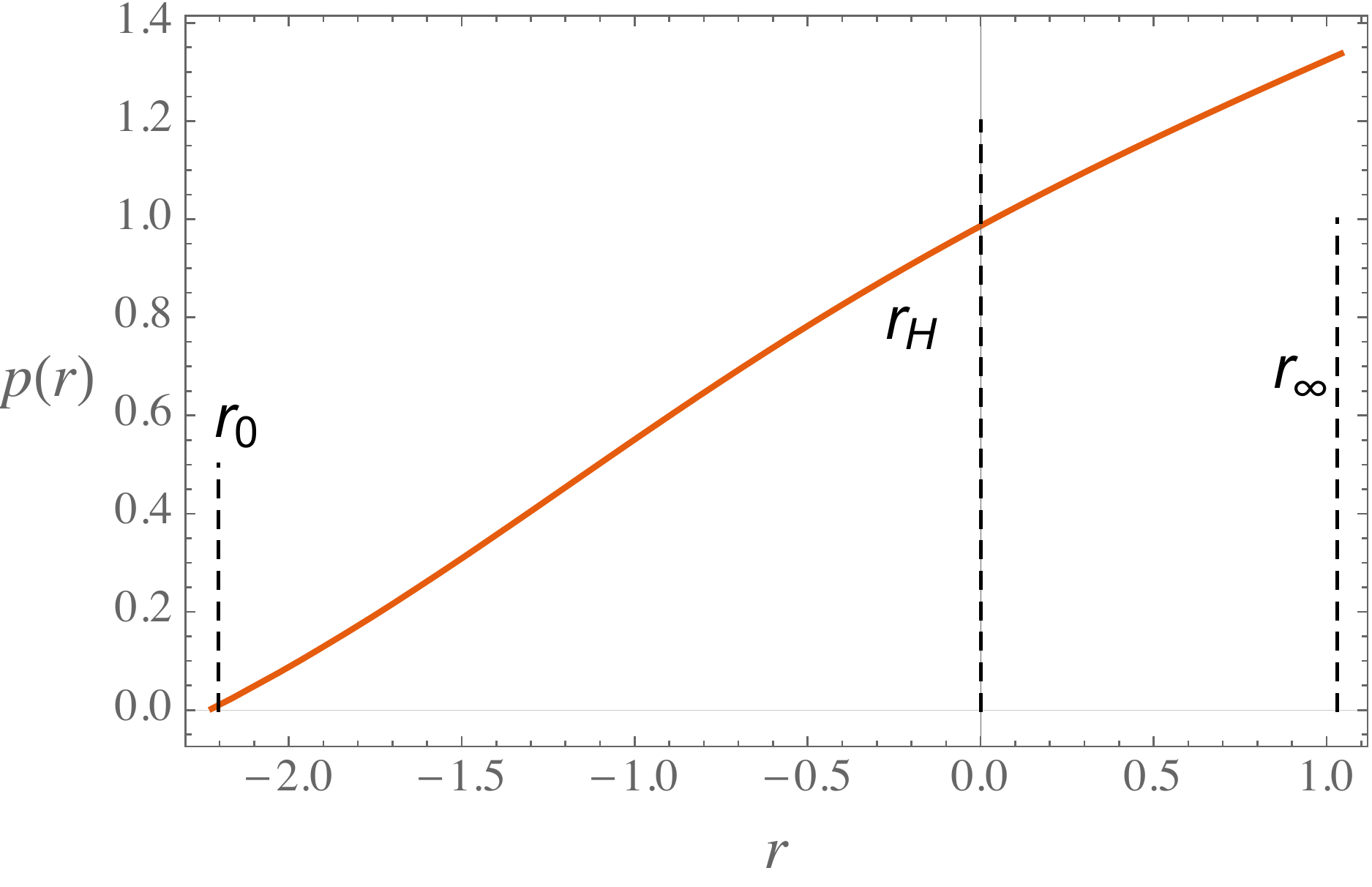}
		\caption{$p(r)$}
		\label{fig:2d-bardeen-p}
	\end{subfigure}
	\begin{subfigure}{.49\textwidth}
	\centering
		\includegraphics[width=\textwidth]{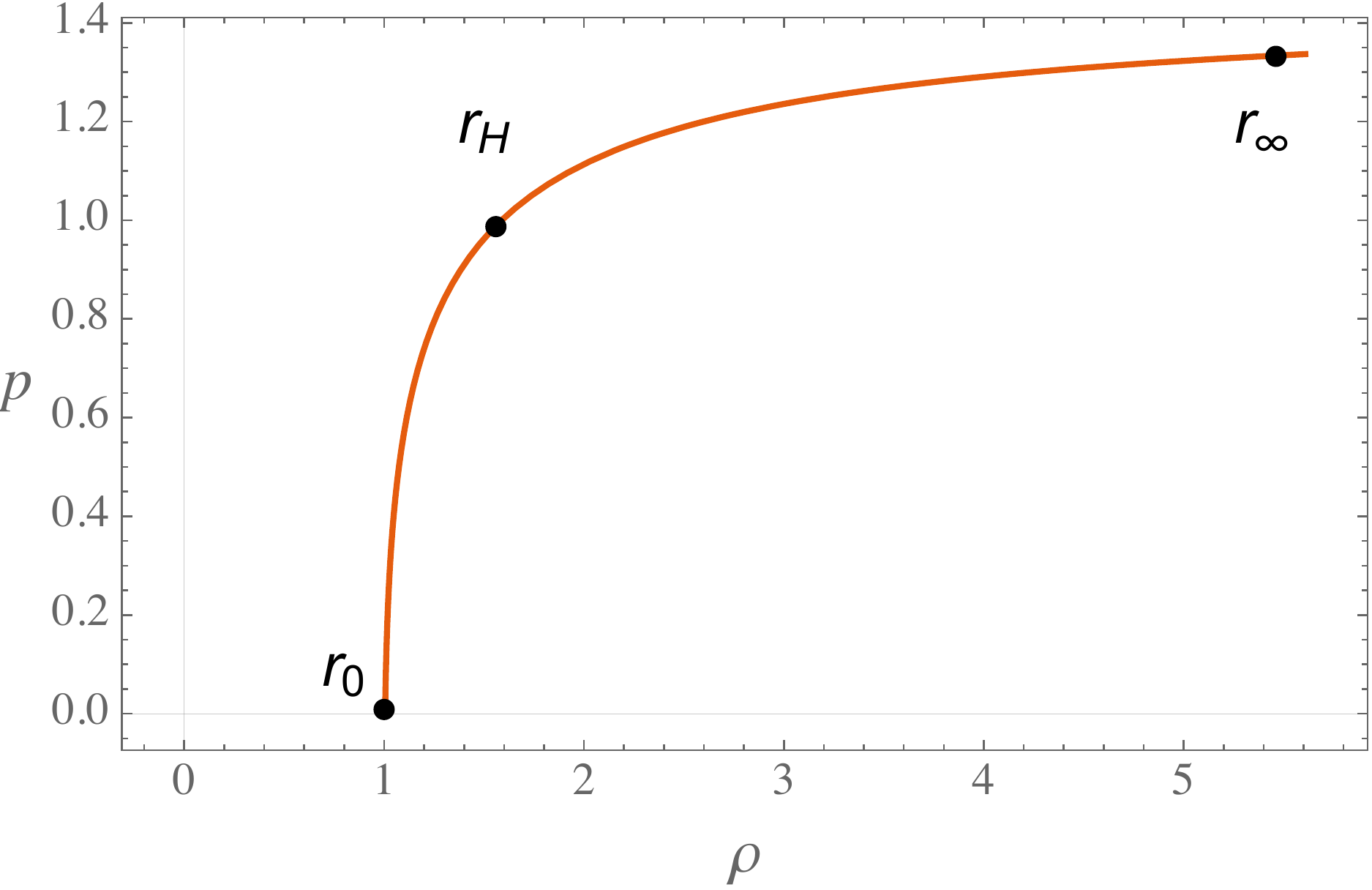}
		\caption{Equation of state}
		\label{fig:2d-bardeen-EoS}
	\end{subfigure}
	\caption{Numerical solutions for the $(2+1)$D repaired Hayward model with $\mu=q=0.5$ and $A=1$, where
	$r_0\approx-2.203$ which is determined by $\xi(r_0)=0$, and $r_\infty\approx 1.030$ which is determined by a very large value of $\xi$.}
	\label{fig:2d-bardeen}
\end{figure}

Similarly, by applying Eq.\ \eqref{eq:phy-var-es}
we find the asymptotic behaviors for the situation with the positive sign around $r=0$,  
\begin{equation}
c^+_0\to \frac{\me^{-r}}{c_{17}},\qquad
v^+_0\to A,\qquad
\rho^+_0\to 1,\qquad
p'^+_0\to 0,
\end{equation}
and in the limit of $r\to\infty$,
\begin{equation}
c^+_\infty\to -\frac{1}{2} k (k r+c_{17}),\qquad
v^+_\infty\to -\frac{A (k r+c_{17})}{2 k},\qquad 
\rho ^+_\infty\to -\frac{2 k}{k r+c_{17}},\qquad
p'^+_\infty\to \frac{k^4}{2}.
\end{equation}
The phenomenon of transonic flows occurs outside the horizon and
the Mach number converges to a constant $\mathcal{M}\to A/k^2$ as $r$ approaches infinity.
The numerical analysis is shown in Fig.\ \ref{fig:2d-bardeen}, where we have adopted the setting $A=1$, $q=1/2$, and $\mu=1/2$, together with the condition $\xi^+(0)=1+\sqrt{2}$. 
The upper boundary of Mach numbers is determined numerically, $\mathcal{M}\le (1+\sqrt{17})/4$.

%%%%%%%%%%%%%%%%%%%%%%%%%%%%%%%%%%%%%%%%%%%%%%%%%%%%%%%%%%%%%%%%%%
\section{Conclusions and outlooks}
\label{sec:concl}
%%%%%%%%%%%%%%%%%%%%%%%%%%%%%%%%%%%%%%%%%%%%%%%%%%%%%%%%%%%%%%%%%%

As an extension of our previous work \cite{Lan:2021ayk},
we analyze two specific questions on RBHs in the present paper: 
The first is how to remedy astronomical RBHs whose DEC is invalid;
and the second is how to simulate realistic RBHs through acoustic gravity. We emphasize that the research strategies of the previous and present works are completely opposite. In the previous work \cite{Lan:2021ayk}, we construct an acoustic metric which is regular at first, then we investigate the energy conditions of the astronomical counterpart of the acoustic RBH.
In the present work, we remedy an astronomical RBH to ensure the DEC and finite curvatures at first, and then we simulate it in a fluid.

The DEC of astronomical RBHs 
occupies a fundamental and decisive position in the research of RBHs.
It determines whether an RBH is observable in the universe or is only theoretical.
On the premise of the DEC,
we have remedied several widely known RBHs whose DEC was broken \cite{Maeda:2021jdc}. In other words, the dominant energy condition, which is violated in the regular black holes listed in Ref. \cite{Maeda:2021jdc}, can be recovered using our strategy given in Sec.\ \ref{sec:remedy}. Thus, these regular black holes revert back to the realistic.
The procedure we proposed in the present paper works for a broad class of RBHs with the broken DEC,
in particular, for those models whose shape functions are rational fraction functions of the radial coordinate.
In addition, we have demonstrated that two types of conformally related RBHs 
can never meet the four energy conditions by proving a no-go theorem.

Although the analogue gravity is widely regarded as a tool of gaining insight into general relativity \cite{Barcelo:2005fc}, the first simulation of Schwarzschild and Reissner-Nordstr\"om black holes was not realized until 2021 \cite{deOliveira:2021edr}.
Prior to this simulation, the acoustic counterparts could not distinguish \cite{Visser:1997ux} those astronomical BHs that differ by a conformal factor, where the differences would be shown in the investigations of desired phenomena from the present point of view. For instance, the quasinormal modes (QNMs) of BHs would be affected~\cite{Chen:2019iuo} by conformal factors.
We hope such an analogue made in a fluid would mimic those astronomic BHs just mentioned.
To this end, starting with realistic RBHs, 
we have constructed their counterparts in acoustic gravity.
Our ultimate goals focus on the guidance on simulation of realistic RBHs in a fluid 
and the possibility to distinguish RBHs from SBHs.
Our guidance on simulation has been illustrated by the equations of state, see 
Figs.\ \ref{fig:Bardeen-EoS}, \ref{fig:Fan-Wang-EoS}, and \ref{fig:2d-bardeen-EoS}, and by the speed of sound and velocity of flow, see Figs.\ \ref{fig:Bardeen-Mach}, \ref{fig:Fan-Wang-Mach}, and \ref{fig:2d-bardeen-Mach}. Moreover, the similarities and differences are listed below when we compare our RBHs with the RN BH \cite{deOliveira:2021edr} in the aspects of velocities and Mach numbers, and in the aspects of densities and pressures for $(3+1)$D models, their equatorial sections, and $(2+1)$D models as follows.

\begin{itemize}
\item Velocity and Mach number in (3+1)D models 

The velocity of flow is divergent at $r=0$ and tends to zero as $r\to \infty$ for the RBHs considered in Sec.\ \ref{sec:simulation}, 
thus the Mach number converges to unity as $r\to 0$ and vanishes as $r\to \infty$. 
For the RN BH \cite{deOliveira:2021edr}, the velocity of flow is finite at $r=0$ and converges to a constant as $r\to \infty$, 
thus the Mach number vanishes at $r=0$ and becomes a finite number as $r\to \infty$.

\item Density and pressure in (3+1)D models 

$\rho(r)$ and $p(r)$ for our RBHs in Sec.\ \ref{sec:simulation} have maximums at the same value of $r$, which is related to the roots of $\xi''(r)=0$.
This property leads to a sharp discontinuity in the plot of EoS, which does not occur for the RN BH. 
Meanwhile, the pressures of our RBHs are negatively infinite at $r=0$,
while they are positively infinite at this point for the RN BH.

\item Velocity, Mach number, density, and pressure in the equatorial sections of $(3+1)$D models

The velocity, Mach number, density, and pressure for the equatorial section of the $(3+1)$D RBH constructed in Sec.\ \ref{sec:cylinder} exhibit similar configurations to those of a RN BH \cite{deOliveira:2021edr}, 
which implies that the regularity of our model does not appear in the simulation of equatorial sections.
Note that we have   
applied the method~\cite{deOliveira:2021edr} to turn a non-compact dimension into a compact one  
in the construction of the relation between $\xi$ and $r$.

\item Velocity and Mach number in $(2+1)$D models 

The velocity and Mach number of our $(2+1)$D RBH constructed in Sec.\ \ref{sec:low-dimension} differ greatly from those of the $(2+1)$D RN BH. At first, our RBH may have only one horizon; then, the transonic flow of our RBH occurs outside the horizon. For the $(2+1)$D RN BH, it has two horizons and its transonic flow occurs between the inner and outer horizons.

\item Density and pressure in $(2+1)$D models

The differences between our $(2+1)$D RBH constructed in Sec.\ \ref{sec:low-dimension} and the $(2+1)$D RN BH \cite{deOliveira:2021edr} are obvious in the density and pressure. For our $(2+1)$D RBH, $\rho(r)$ and $p(r)$ are no longer divergent at $r=0$ and 
the EoS becomes smooth in the whole region of $r$, see Figs.~\ref{fig:2d-bardeen-rho}, \ref{fig:2d-bardeen-p}, and \ref{fig:2d-bardeen-EoS}. 
Meanwhile, the behaviors of the two variables in the $(2+1)$D RBH differ from those in the equatorial sections of $(3+1)$D models, despite the fact that we use the identical compactification in both situations. This difference indicates that the dimensions of RBHs affect the properties of flow simulations.
\end{itemize}

Finally, we summarize that the EoS of fluid can be used to simulate
realistic RBHs. Meanwhile, we show
that the acoustic analogues of RBHs have apparently different features
from those of SBHs, such as the RN BH, and that 
the differences are indeed caused by singularities.
In other words, the acoustic gravity can be applied as a tool to study astronomic RBHs, 
which also offers a theoretical basis to investigate more phenomena of astronomic RBHs in a fluid.

\section*{Acknowledgments}
This work was supported in part by the National Natural Science Foundation of China under Grant Nos.\ 11675081 and 12175108.

\appendix

%%%%%%%%%%%%%%%%%%%%%%%%%%%%%%%%%%%%%%%%%%%%%%%%%%%%%%
\section{The differential inequalities}
\label{app:diff-ineq}
%%%%%%%%%%%%%%%%%%%%%%%%%%%%%%%%%%%%%%%%%%%%%%%%%%%%%%

Here we solve the  differential inequalities appeared in this paper, 
in particular, in Sec.\ \ref{sec:correct-RBHs}.
We start with solving the differential inequality, $\xi \sigma''\le 2 \sigma'$, 
with the boundary conditions, $\sigma(0)=0=\sigma'(0)$.
It can be rewritten as
\begin{equation}
    \frac{\dif}{\dif \xi} \left(3 \sigma -\xi \sigma '\right)\ge 0.
\end{equation}
After considering the boundary conditions, we obtain $3 \sigma -\xi \sigma '\ge 0$. 
Next, multiplying $\xi^{-4}$ on its both sides, we derive 
\begin{equation}
    3 \xi^{-4} \sigma-
    \xi^{-3}\sigma'=\frac{\dif}{\dif \xi}\left(
    -\xi^{-3} \sigma
    \right)\ge 0.
\end{equation}
If we define $\sigma_0:=\lim_{\xi\to 0} \sigma/\xi^3$,
we arrive at the solution,
\begin{equation}
    \sigma\le \sigma_0 \xi^3,
\end{equation}
which can also be obtained when one directly uses the differential form of the 
Gr\"onwall-Bellman lemma \cite{Demidovich:2008ms}.

Similarly, we can obtain $\sigma\ge 0$ from $\sigma'\ge 0$ when the boundary condition, $\sigma(0)=0$, is considered.

We emphasize that the differential inequalities' solutions provide the necessary condition for a RBH to meet the energy conditions, but not the sufficient one. For example,
taking a bell-shaped function, 
$$
\sigma=\frac{1}{\left(\xi-\frac{1}{2\xi}\right)^4+1},
$$
we can see that the inequality's solution, $\sigma\ge 0$, is satisfied
but the WEC and DEC are violated
because of $\sigma'\ngeq 0$.
However, if $\sigma\ge 0$ is broken, then the WEC and DEC must be violated.  
In addition, the Hayward BH gives us another example, that is, it satisfies $0\le\sigma\le \sigma_0 \xi^3$, 
but breaks the DEC. 
The reason comes from the characteristics of differential inequalities,
i.e., a differential inequality signifies 
that all functions satisfying this differential inequality must be bounded by its solution,
while the functions bounded by the solution may not necessarily meet the original differential inequality.

%%%%%%%%%%%%%%%%%%%%%%%%%%%%
\section{Local properties of the differential inequalities}
\label{app:local-diff-ineq}
%%%%%%%%%%%%%%%%%%%%%%%%%%%%%%%%%%%%%

Now we give an explanation from the energy conditions by analyzing the local properties of a realistic RBH at its center, 
that is, why a realistic RBH cannot have a flat or an AdS core around its center.
The similar discussion can be found  in Ref.\ \cite{Maeda:2021jdc}.

Summarizing the energy conditions Eq.\ \eqref{eq:ECDs}, 
we have four differential inequalities in total,
\begin{equation}
\label{eq:diff-ineq-ec}
\sigma '\ge 0,\qquad
2 \sigma '-\xi \sigma ''\ge 0,\qquad 
2 \sigma '+\xi \sigma ''\ge 0,\qquad
\sigma ''\le 0.
\end{equation}
Supposing $\xi\ll 1$, we expand $\sigma$  by an asymptotic series,
\begin{equation}
    \sigma=\xi^3\sum _{n=0}^{\infty } a_n \xi^{n},
\end{equation}
meanwhile, we have the property, $\abs{a_n \xi^n}\gg\abs{a_{n+1} \xi^{n+1}}$, as $\xi$ approaches $0$.
Then substituting the above series into Eq.\ \eqref{eq:diff-ineq-ec}, we obtain 
\begin{subequations}
\begin{align}
    \sigma '\ge 0 &: \;\;\;\;\; \sum _{n=0}^{\infty } (n+3) a_n \xi^{n+2}\ge 0,\label{eq:1st-inq}
                     \\
    2 \sigma '\ge \xi \sigma '' &: \;\;\;\;\;  \sum _{n=1}^{\infty } n (n+3) a_n \xi^{n+2}\le 0, \label{eq:2nd-inq}\\ 
     -2 \sigma '\le \xi \sigma ''&: \;\;\;\;\;  \sum _{n=0}^{\infty } (n+3) (n+4) a_n \xi^{n+2}\ge 0,\label{eq:3rd-inq}
                         \\
     \sigma ''\le 0 &: \;\;\;\;\;  \sum _{n=0}^{\infty } (n+2) (n+3) a_n \xi^{n+1}\le 0.\label{eq:4th-inq}
     \end{align}
\end{subequations}
If $a_0\neq 0$ and $a_1\neq 0$,
the leading terms of Eqs.\ \eqref{eq:1st-inq} and \eqref{eq:3rd-inq} lead to $a_0> 0$,
which inevitably  violates Eq.\ \eqref{eq:4th-inq}, 
and the leading term of Eq.\ \eqref{eq:2nd-inq} gives $a_1< 0$.
If $a_0= 0$, the DEC must be broken because Eq.\ \eqref{eq:1st-inq} and Eq.\ \eqref{eq:2nd-inq} lead to the results that are contradictory to each other. 
If $a_0\neq 0$ and $a_1= 0$, the leading term of Eq.\ \eqref{eq:2nd-inq} gives $a_m< 0$, $m>1$, 
where $m$ is the ordinal  number of the first non-zero term.

In contrast, based on Ref.~\cite{Lan:2021ngq} we know 
\begin{equation}
\label{eq:series-sigma}
\sigma = \xi^3 \sum _{n=0}^{\infty } \frac{\xi^n R^{(n)}(0)}{2 M (n+3) (n+4) n!}
\sim \frac{\xi^3 R(0)}{24 M}+\frac{\xi^4 R'(0)}{40 M}+O(\xi^5),
\end{equation}
as $\xi$ approaches $0$.
Thus the case of $a_0> 0$ and $a_1< 0$ implies 
\begin{equation}
    R(0)> 0
    \quad \text{and}\quad
    R'(0)< 0.
\end{equation}
For $a_1=0$ but $a_2\neq 0$, we obtain $R(0)> 0$ and $R''(0)< 0$, 
i.e. $\xi=0$ is a local maximum.
Moreover, if an RBH is Ricci flat at its center, $R(0)=0$, or it has an AdS core, $R(0)< 0$, its energy conditions must be violated around the center.

%%%%%%%%%%%%%%%%%%%%%%%%%%%%%%
\section{The derivation of Eq.\ \eqref{eq:dec-generic}}
\label{app:dec-generic}
%%%%%%%%%%%%%%%%%%%%%%%%%%%%%%

In order to derive the conditions given by Eq.\ \eqref{eq:dec-generic}, we substitute Eq.\ \eqref{eq:generic-sigma} into the
DEC of Eq.\ \eqref{eq:ECDs}. Since the DEC holds for $\xi\in[0,\infty)$ and $q\in[0,\infty)$, we obtain
\begin{equation}
3-\mu  \nu \geq 0,\qquad
(\mu  \nu -4) (\mu  \nu -3)\ge 0,\qquad
24-\mu  (\mu +7) \nu \ge 0.\label{104}
\end{equation}
Meanwhile, the asymptotic flatness demands
\begin{equation}
3-\mu  \nu <1.\label{105}
\end{equation}
Then, combining Eqs.~(\ref{104}) and (\ref{105}) and considering the positiveness of all parameters, we finally arrive at Eq.\ \eqref{eq:dec-generic}.

%%%%%%%%%%%%%%%%%%%%%%%%%%%%%%%%%%%%%%%%%%%%%%%%%%%%%%%%%%%%%%%%%%
\section{The $d$-dimensional regular black holes}
\label{app:high-rbh}
%%%%%%%%%%%%%%%%%%%%%%%%%%%%%%%%%%%%%%%%%%%%%%%%%%%%%%%%%%%%%%%%%

Now we derive the regularity conditions which have been applied in Sec.\ \ref{sec:low-dimension}.
We write down the $d$-dimensional metric with the spherical symmetry \cite{Emparan:2008eg},
\begin{equation}
\label{eq:d-metric}
\dif s^2= -f \dif t^2 +f^{-1} \dif \xi^2 +\xi^2 \dif \Omega_{d-2}^2,
\end{equation}
where $f$ is shape function,  $f=1-\mu\sigma(\xi)/\xi^{d-3}$, and $\mu$ is mass-like parameter.
As we did in Ref.\ \cite{Lan:2021ngq}, we are going to use the following three curvatures, 
\begin{eqnarray}
R & =&\frac{\mu }{\xi^{d-2}}\left(2 \sigma '+\xi \sigma ''\right),\\
W&=&\frac{(d-3) \mu^2}{(d-1) \xi^{2 d-2}}
	\left[(d-2) (d-1) \sigma
	-2 (d-2) \xi \sigma ' 
	+\xi^2 \sigma ''\right]^2,\\
E&=&\frac{2  \mu^2 }{d \xi^{2 d-4}}\left[(d-2) \sigma '-\xi \sigma ''\right]^2,
\end{eqnarray}
to represent $\sigma$ and its derivatives.
\iffalse
\begin{equation}
\begin{split}
R &=\frac{2 M }{r^3}
\left[2 (d-3) r \sigma '+(d-4) (d-3) \sigma+r^2 \sigma ''\right]
\\
 W &= 
 \frac{4 (d-3) }{(d-1) }\frac{M^2 }{r^6}
\left(6 \sigma-4 r\sigma '+ r^2 \sigma ''\right)^2\\
E&=
\frac{8 M^2 }{d r^6}
\left[r (d-6) \sigma '+r^2 \sigma ''-3 (d-4) \sigma \right]^2
 \end{split}
\end{equation}
\fi
With the help of the relations,
\begin{equation}
W=K-\frac{4R_2}{d-2} +\frac{2R^2}{(d-1)(d-2)} 
,\qquad
E=\frac{4 R_2}{d-2}-\frac{4 R^2}{d(d-2)},
\end{equation}
we arrive at
\begin{eqnarray}
\sigma &=&\frac{\xi^{d-1}}{(d-2) (d-1) d \mu}
 \left[
 	(d-2) R+\mathfrak{s}_2 (d-1) \sqrt{2 d E} +\mathfrak{s}_1 d  \sqrt{\frac{d-1}{d-3} W}
\right],
\\
\sigma ' &=&\frac{\xi^{d-2} }{2 d \mu}\left(2 R+\mathfrak{s}_2\sqrt{2 d E} \right),\\
\sigma '' &=& \frac{\xi^{d-3} }{d \mu}\left[(d-2) R-\mathfrak{s}_2\sqrt{2d E} \right],
\end{eqnarray}
\iffalse
\begin{equation}
\begin{split}
\sigma =
\frac{r^3 }{2 (d-2) (d-1) d M}
\Big[(d-2) R
	-&\mathfrak{s}_1(d-1)\sqrt{2d E}   
	+ \mathfrak{s}_2 d\sqrt{\frac{d-1}{d-3}W}\Big]
\\
\sigma '=
\frac{r^2}{4 (d-2) (d-1) d M}
 \Big[6 (d-2) R
 	+&\mathfrak{s}_1 (d-6) (d-1) \sqrt{2d E}- \\
 	&-\mathfrak{s}_2 2 (d-4) d \sqrt{\frac{(d-1) W}{d-3}}\Big]
\\
\sigma ''= 
\frac{r }{2 (d-2) (d-1) d M}
\Big[6 (d-2) R
	+&\mathfrak{s}_1 2(d-3) (d-1) \sqrt{2d E}+  \\
	&+\mathfrak{s}_2(d-4) (d-3) d  \sqrt{\frac{(d-1) W}{d-3}}\Big]
 \end{split}
\end{equation}
\fi
where $\mathfrak{s}_{1,2}=\pm$ are two signs which are not much important for the discussion of the finiteness of curvatures.
Meanwhile, it is not difficult to see that the metric Eq.\ \eqref{eq:d-metric} has finite curvatures if $\sigma$ has asymptotic relation $\sigma\lesssim O(\xi^{d-1})$.

%%%%%%%%%%%%%%%%%%%%%%%%%%%%%%%%%%%%%%%%%%%%%%%%%%%%%%%%%%
\section{Asymptotic solutions of the differential equation}
\label{app:asym-sol}
%%%%%%%%%%%%%%%%%%%%%%%%%%%%%%%%%%%%%%%%%%%%%%%%%%%%%%%%%%

We analyze the local properties of the differential equation Eq.\ \eqref{eq:master-relation},
\begin{equation}
A^2 \xi ^4 r'(\xi )^4+F(\xi) \xi ^2 r(\xi )^6 r'(\xi )^2-r(\xi )^8=0,
\end{equation}
at the two boundaries, $\xi \to 0$ and $\xi \to \infty$, by means of the dominant balance \cite{Bender:2013amm,Paulsen:2013ap}.
Here $\xi\in [0,\infty)$ is the radial coordinate of RBHs, while $r$ is radial coordinate of fluids with the spherical symmetry.

First of all, we consider the asymptoticity of the shape function as $\xi\to 0$, 
$$
F\sim F_0\coloneqq 1-\frac{R(0)}{12} \xi^2.
$$
According to the dominant balance, 
we can separate the discussions into three situations.

In the first case, we have
\begin{subequations}
\begin{equation}
\label{eq:zero-case1}
A^2 \xi ^4 r'(\xi )^4\sim -F_0 \xi ^2 r(\xi )^6 r'(\xi )^2,
\end{equation}
\begin{equation}
A^2 \xi ^4 r'(\xi )^4\gg r(\xi )^8,\qquad
F_0 \xi ^2 r(\xi )^6 r'(\xi )^2\gg r(\xi )^8,
\end{equation}
\end{subequations}
where Eq.\ \eqref{eq:zero-case1} gives the solution,
\begin{equation}
r_{\pm}^{-2} =
-\frac{2 }{A} \sqrt{-1+ \frac{R(0)}{12}\xi ^2}
+\frac{2 }{A} \tan^{-1}\left[\sqrt{-1+ \frac{R(0)}{12}\xi ^2}\right]
-2 {\tilde c}_1,
\end{equation}
and ${\tilde c}_1$ is an integration constant. This solution becomes complex as $\xi \to 0$, which contradicts the physical requirement that $r(\xi)$ must be real.

In the second case, the asymptotic relations become
\begin{subequations}
\begin{equation}
\label{eq:zero-case2}
F_0 \xi ^2 r(\xi )^6 r'(\xi )^2\sim r(\xi )^8,
\end{equation}
\begin{equation}
\label{eq:zero-assump-case2}
F_0 \xi ^2 r(\xi )^6 r'(\xi )^2\gg 
A^2 \xi ^4 r'(\xi )^4,\qquad
r(\xi )^8\gg A^2 \xi ^4 r'(\xi )^4.
\end{equation}
\end{subequations}
The solution of Eq.\ \eqref{eq:zero-case2} is
\begin{equation}
r_\pm = \frac{ \xi  \sqrt{3R(0)}}{6\pm 6 \sqrt{1- \frac{R(0)}{12}\xi ^2}}.
\end{equation}
However, $r_+$ is not consistent with the asymptotic assumption depicted by Eq.\ \eqref{eq:zero-assump-case2}.
For $R(0)\neq 0$, we have a divergent limit as $\xi\to 0$,
\begin{equation}
 \lim_{\xi\to0} \frac{A^2 \xi ^4 r_+'(\xi )^4}{F_0 \xi ^2 r_+(\xi )^6 r_+'(\xi )^2}\to\infty.
\end{equation}
As to $r_-$, we note that it becomes complex if $R(0)< 0$, thus this second case should also be ruled out.

In the third case, we suppose
\begin{subequations}
\begin{equation}
\label{eq:zero-case3}
A^2 \xi ^4 r'(\xi )^4\sim r(\xi )^8,
\end{equation}
\begin{equation}
\label{eq:zero-assump-case3}
A^2 \xi ^4 r'(\xi )^4\gg F_0 \xi ^2 r(\xi )^6 r'(\xi )^2,\quad
r(\xi )^8\gg F_0 \xi ^2 r(\xi )^6 r'(\xi )^2.
\end{equation}
\end{subequations}
Eq.\ \eqref{eq:zero-case3} provides the solution,
\begin{equation}
\label{eq:zero-sol-case3}
r_\pm =-\frac{\sqrt{A}}{{\tilde c}_2\pm\ln (\xi )},
\quad \text{or}\qquad
\xi =  {\tilde c}_3\exp\left(\mp
\frac{\sqrt{A}}{r_\pm}
\right),
\end{equation}
where ${\tilde c}_2$ and ${\tilde c}_3$ are integration constants. This solution is consistent with the the asymptotic assumption Eq.\ \eqref{eq:zero-assump-case3} because we have
\begin{equation}
\lim_{\xi\to0} \frac{F_0 \xi ^2 r(\xi )^6 r'(\xi )^2}{A^2 \xi ^4 r'(\xi )^4}
=\lim_{\xi\to0} \frac{1-\frac{R(0)}{12} \xi ^2}{\left[{\tilde c}_2\pm \ln (\xi ) \right]^2}=0.
\end{equation}
However, we exclude $r_-$ from physical solutions due to $r_-<0$ when $\xi\sim 0^+$.

Next, we turn to the asymptotic solutions as $\xi\to \infty$. The shape function then becomes
$$
F\sim F_\infty\coloneqq 1-2M \xi^{-n},
$$
where $0<n \le 1$.
Similarly, the discussion can also be separated into three situations.

In the first case, we have
\begin{subequations}
\begin{equation}
\label{eq:inf-case1}
\xi ^2 F_\infty r(\xi )^6 r'(\xi )^2\sim r(\xi )^8,
\end{equation}
\begin{equation}
\label{eq:inf-assump-case1}
\xi ^2 F_\infty r(\xi )^6 r'(\xi )^2\gg A^2 \xi ^4 r'(\xi )^4,
\qquad
 r(\xi )^8\gg A^2 \xi ^4 r'(\xi )^4.
\end{equation}
\end{subequations}
The asymptotic assumption Eq.\ \eqref{eq:inf-case1} gives two solutions,
\begin{equation}
r_\pm={\tilde c}_4 \left(\sqrt{\xi ^{n}/(2M)}+\sqrt{\xi ^n/(2M)-1}\right)^{\pm 2/n}\sim 
 {\tilde c}_5  \xi^{\pm1},
\end{equation}
where ${\tilde c}_4$ and ${\tilde c}_5$ are integration constants.
It can be verified that $r_-$ contradicts to the asymptotic assumption Eq.\ \eqref{eq:inf-assump-case1},
while $r_+$ does not. 

In the second case, we suppose
\begin{subequations}
\begin{equation}
A^2 \xi ^4 r'(\xi )^4\sim r(\xi )^8,
\end{equation}
\begin{equation}
A^2 \xi ^4 r'(\xi )^4\gg \xi ^2 r(\xi )^6 F_\infty r'(\xi )^2,\qquad
r(\xi )^8\gg \xi ^2 r(\xi )^6 F_\infty r'(\xi )^2.
\end{equation}
\end{subequations}
The first relation leads to 
\begin{equation}
r_\pm =-\frac{ \sqrt{A}}{{\tilde c}_6\pm\ln (\xi )},
\end{equation}
where ${\tilde c}_6$ is an integration constant. 
We can see that $r_\pm$ converge to zero as $\xi\to\infty$. 
This implies that the transformation between $r$ and $\xi$ is not injective, 
which is obvious because both $\xi=0$ and  $\xi=\infty$ map to the single point $r=0$.
As a result, we eliminate this case.

In the last case, the asymptotic assumption involves
\begin{subequations}
\begin{equation}
A^2 \xi ^4 r'(\xi )^4\sim -\xi ^2 r(\xi )^6 F_\infty r'(\xi )^2,
\end{equation}
\begin{equation}
A^2 \xi ^4 r'(\xi )^4\gg r(\xi )^8,\quad
\xi ^2 r(\xi )^6 F_\infty r'(\xi )^2\gg r(\xi )^8,
\end{equation}
\end{subequations}
whose solutions are inevitably complex. Thus, this case is not in our consideration.

\bibliographystyle{utphys}
\bibliography{references}

\end{document}